\documentclass[10pt,aps,prc,amsmath,amssymb,floatfix,twocolumn,nofootinbib,superscriptaddress]{revtex4-2}

\usepackage{footmisc}
\usepackage{tabularx}
\usepackage[utf8]{inputenc}
\usepackage[T1]{fontenc}
\usepackage{graphicx}
\usepackage{isotope}
\usepackage{mathtools}
\usepackage{physics}
\usepackage[pdfpagelabels, pdfencoding=auto, psdextra]{hyperref}
\usepackage{xparse}
\hypersetup{%
 pdfsubject={PMM-IMSRG emulator for the nuclear equation of state with quantified uncertainties},
 pdfkeywords={nuclear physics},
 unicode = true,
 breaklinks = true,
 colorlinks = true,
 linkcolor = blue,
 menucolor = blue,
 citecolor = blue,
 urlcolor = blue
}
\usepackage[inkscapelatex=false]{svg}
\usepackage{braket}
\usepackage{siunitx}
\usepackage{mleftright}
\usepackage{placeins}
\usepackage{aas_macros}

\DeclareMathOperator*{\arginf}{arg\,inf}

\DeclareMathOperator*{\mean}{mean}
\DeclareMathOperator*{\clipfunc}{clip}

\newcommand{\ntwlo}{\ensuremath{\text{N}^2\text{LO}}}
\newcommand{\nthlo}{\ensuremath{\text{N}^3\text{LO}}}
\newcommand{\bc}{\ensuremath{\mathbf{c}}}

\newcommand{\comment}[1]{}

\newcommand{\refket}{|\Phi \rangle}
\newcommand{\refbra}{\langle \Phi |}

\newcommand{\orcid}[1]{\href{https://orcid.org/#1}{\includegraphics[scale=0.055]{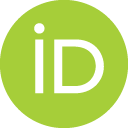}}}

\newcommand{\cd}{\ensuremath{c_D}}
\newcommand{\ce}{\ensuremath{c_E}}
\newcommand{\dt}{\ensuremath{D_2}}
\newcommand{\ft}{\ensuremath{F_2}}
\newcommand{\unc}[1]{\ensuremath{\mathcal{U}\mleft(#1\mright)}}

\newcommand{\fm}{\ensuremath{\si{\femto\meter}}}

\newcommand{\fmiq}{\ensuremath{\si{\per\cubic\femto\meter}}}

\newcommand{\MeV}{\ensuremath{\si{\mega\eV}}}

\DeclareSIUnit[quantity-product = \,]\flop{\text{FLOP}}

\newcommand{\bigO}[1]{\ensuremath{\mathcal{O}\left(#1\right)}}

\newcommand{\pmmimsrg}{PMM--IMSRG}

\begin{document}

\title{\pmmimsrg{} emulator for the nuclear equation of state with quantified uncertainties}

\author{Patrick Cook~\orcid{0000-0002-7934-5428}}
\email{cookpat4@msu.edu}
\thanks{These authors contributed equally.}
\affiliation{\href{https://ror.org/03r4g9w46}{Facility for Rare Isotope Beams}, \href{https://ror.org/05hs6h993}{Michigan State University}, East Lansing, MI~48824, USA}
\affiliation{Department of Physics and Astronomy, \href{https://ror.org/05hs6h993}{Michigan State University}, East Lansing, MI~48824, USA}

\author{Kang Yu~\orcid{0009-0005-1099-7697}}
\email{yuk@frib.msu.edu}
\thanks{These authors contributed equally.}
\affiliation{\href{https://ror.org/03r4g9w46}{Facility for Rare Isotope Beams}, \href{https://ror.org/05hs6h993}{Michigan State University}, East Lansing, MI~48824, USA}
\affiliation{Department of Physics and Astronomy, \href{https://ror.org/05hs6h993}{Michigan State University}, East Lansing, MI~48824, USA}

\author{Christian Drischler~\orcid{0000-0003-1534-6285}}
\email{drischler@ohio.edu}
\affiliation{Department of Physics and Astronomy, \href{https://ror.org/01jr3y717}{Ohio University}, Athens, OH~45701, USA}
\affiliation{\href{https://ror.org/03r4g9w46}{Facility for Rare Isotope Beams}, \href{https://ror.org/05hs6h993}{Michigan State University}, East Lansing, MI~48824, USA}

\author{Scott K.\ Bogner~\orcid{0000-0003-1584-011X}}
\email{bogner@frib.msu.edu}
\affiliation{\href{https://ror.org/03r4g9w46}{Facility for Rare Isotope Beams}, \href{https://ror.org/05hs6h993}{Michigan State University}, East Lansing, MI~48824, USA}
\affiliation{Department of Physics and Astronomy, \href{https://ror.org/05hs6h993}{Michigan State University}, East Lansing, MI~48824, USA}

\begin{abstract}

We introduce a hybrid emulator for in-medium similarity renormalization group (IMSRG) calculations of nuclear matter, based on chiral nucleon-nucleon and three-nucleon interactions and an implicit-reduced-basis method emulator constructed from parametric matrix models (PMMs) which is capable of rigorously estimating its uncertainties via conformal predictions.
The resulting PMM--IMSRG emulator enables fast and accurate predictions with trustworthy confidence intervals of the nuclear equation of state (EOS) across a wide range of input parameters, including low-energy couplings, IMSRG flow parameters, densities, and basis sizes.
This framework provides the foundation for principled uncertainty quantification of the nuclear EOS and enables computationally demanding applications such as Bayesian parameter estimation using our IMSRG calculations.
As a first application, we present results for the coupling constants of the two quark-mass-dependent three-nucleon interactions, recently identified to contribute at next-to-next-to-leading order in the chiral expansion based on a renormalization-group analysis, by fitting them to empirical saturation properties.
We then propagate both parametric and emulator uncertainties to the EOS in the limits of pure neutron matter and symmetric nuclear matter.

\end{abstract}

\maketitle

\section{Introduction}
\label{sec:intro}

The nuclear equation of state (EOS) plays a central role in understanding dense astrophysical objects such as neutron stars, providing a crucial bridge between microscopic nuclear physics and astrophysical observables~\cite{Drischler:2021kxf,Sorensen:2023zkk,MUSES:2023hyz,Chatziioannou:2024jsr,Agarwal:2025ezo}.
Considerable progress has been made in constraining the EOS microscopically, driven by advances in chiral effective field theory (EFT) and the development of sophisticated many-body methods for nuclear matter, including
many-body perturbation theory~\cite{Drischler:2026vdm,Keller:2020qhx,Keller:2022crb,Holt:2016pjb,Wen:2024shw},
diagrammatic Monte Carlo~\cite{Brolli:2025ehf},
coupled-cluster theory~\cite{Hagen:2013yba,Baardsen:2013vwa,Jiang:2022tzf},
quantum Monte Carlo techniques~\cite{Lonardoni:2019ypg,Arthuis:2022ixv,Fore:2024exa,Lovato:2026erx, Hu:2025udi}, and
the self-consistent Green’s functions method~\cite{FrancescoMarino:2026dzw,Dickhoff:2004xx, Rios:2020oad}.
In particular, recent developments have established an IMSRG framework for computing the nuclear EOS, which provides the foundation for the present work~\cite{Udiani_thesis,Udiani_IMSRG} (see also Ref.~\cite{Zhen:2025gfy}).

However, repeated predictions of the nuclear EOS pose significant computational challenges, e.g., when sampling the parameter spaces of the underlying interactions for principled uncertainty quantification (UQ), because these \textit{ab~initio} many-body calculations are often prohibitively expensive.
Emulators provide an efficient means to overcome these computational challenges by constructing fast and accurate surrogate models for high-fidelity \textit{ab~initio} calculations.
See Refs.~\cite{Duguet:2023wuh,Drischler:2022ipa} for recent review articles and Refs.~\cite{Armstrong:2026aap,Armstrong:2025tza,Somasundaram:2024zse,Jiang:2022oba} for recent applications of nuclear matter emulators.

In this work, we develop a hybrid emulator using a parametric matrix model (PMM) approach~\cite{Cook2025} for IMSRG nuclear matter calculations, designed to embed as much relevant physics as possible while learning the remaining unknown or inaccessible structure from data. Unlike purely data-driven approaches, PMMs preserve the operator structure of the many-body problem by learning low-dimensional representations of the Hamiltonian itself. This feature enables the PMM to retain important physical properties dictated by the structure of the underlying Hamiltonian, while providing a natural route to emulate across model parameters.
Compared to eigenvector continuation~\cite{Frame:2017fah} and other intrusive reduced-basis emulators~\cite{Melendez:2022kid,Drischler:2022ipa}, which rely on explicitly constructed subspaces from high-fidelity training states, PMMs implicitly parameterize the relevant low-dimensional subspaces through learned operators. Using this approach avoids, e.g., the need to store or manipulate large many-body wavefunctions, while still capturing the dominant parametric dependence of the ground-state energy. These features make PMMs particularly well-suited for IMSRG calculations, in which the underlying physics is governed by smoothly varying operators, yet direct evaluation remains computationally expensive.

The resulting \pmmimsrg{} emulator enables fast and accurate predictions of the EOS across a wide range of physical and computational inputs, including chiral EFT low-energy constants, IMSRG flow parameters, densities, and basis sizes.
This capability establishes the foundation for principled UQ of the nuclear EOS and opens the door to a variety of computationally demanding applications.
To demonstrate the efficacy of this framework, we present exploratory results for the coupling constants of the two quark-mass-dependent three-nucleon interactions ($D_2$, $F_2$)~\cite{Vernik:2025czp}, recently identified to contribute at next-to-next-to-leading order (N$^2$LO) in the chiral expansion based on a renormalization-group analysis~\cite{Cirigliano2025Chiral}, by fitting them to empirical saturation properties.
We then propagate both parametric and emulator uncertainties to the EOS in the limits of pure neutron matter (PNM) and symmetric nuclear matter (SNM).

\begin{table}[tb]
\renewcommand{\arraystretch}{1.3}
\setlength{\tabcolsep}{6pt}
\caption{
Notation used in this work.
}
\label{tab:notation}
\begin{tabularx}{\linewidth}{lX}
\hline
\hline
Notation & Description \\
\colrule
$\nu$ & chiral order of the calculation, i.e., $\nu=3$ for \ntwlo{}\\
$s$ & IMSRG flow parameter \\
$\bc$ & vector of LECs, with components $c_i$ \\
$N_s$ & number of allowed single particle states\\
$Y_p$ & proton fraction \\
$n$ & nucleon density \\
$E_0$ & (ground-state) energy per particle\\
$n_\text{sat}$, $E_\text{sat}$ & saturation point in SNM\\
$X_H$ & $\equiv \left[\bc,\nu,N_s,Y_p,n\right]$ \\
$P_1$ & projector onto reduced in-medium Hamiltonian space\\
$P_2$ & projector onto reduced free space Hamiltonian space\\
$f_i(x_i), f_j(N_s, Y_p, n)$ & data-driven hidden features for IMSRG basis\\
$\alpha(X_H), \beta(X_H)$ & data-driven hidden features for IMSRG flow dependence\\
$X_{\text{SNM}}$ or $X_{\text{PNM}}$ & $\equiv\left[\bc,\,\nu,\,N_s,\,Y_p =0.5~\text{or}~0.0,\,n\right]$ \\
$c_D, c_E, \dt, \ft$ & LECs considered in this work\\
$\mathcal{T}$ & training dataset\\
$\alpha$ & miscoverage interval\\
\hline
\hline
\end{tabularx}
\end{table}

The remainder of this paper is organized as follows.
In Sec.~\ref{sec:imsrg}, we describe our IMSRG framework for nuclear matter calculations with nucleon-nucleon (NN) and three-nucleon (3N) interactions at the normal-ordered two-body level.
Section~\ref{sec:method} introduces the PMM emulation framework for IMSRG calculations of the nuclear EOS, including the general formalism, strategies for emulation across chiral EFT low-energy constants (LECs), IMSRG flow parameters, densities, and basis sizes, as well as the associated UQ.
In Sec.~\ref{sec:results}, we present proof-of-principle applications of \pmmimsrg.
We discuss the emulator's efficacy, constrain quark-mass-dependent chiral 3N forces at \ntwlo{} in the chiral expansion, and propagate parametric uncertainties to the nuclear EOSs.
Finally, Sec.~\ref{sec:summary} provides a summary and outlook.
We use natural units in which $\hbar = c = 1$ and adopt the notation summarized in Table~\ref{tab:notation}.

\section{IMSRG for Nuclear Matter}
\label{sec:imsrg}

In this section, we briefly summarize the formalism and implementation of the IMSRG for our nuclear matter calculations. We refer the reader to Refs.~\cite{Hergert:2016jk,Hergert2017,Hergert:2016etg,Stroberg:2019mxo} for comprehensive review articles on the IMSRG, and Refs.~\cite{Udiani_thesis,Udiani_IMSRG,Zhen:2025gfy} for details on IMSRG calculations of nuclear matter.
\subsection{Discrete momentum basis}
\label{subsec:spbasis}
We model nuclear matter as a system of $A$ nucleons confined to a cubic box of length $L$ with periodic boundary conditions. The single-particle basis states are then momentum eigenstates
\begin{equation}
|p\rangle \equiv |{\bf k}_p \sigma_p \tau_p\rangle\,,
\end{equation}
where $\sigma_p$ and $\tau_p$ denote the spin and isospin projections, and the momentum components are quantized as
\begin{equation}
k_{i} = \frac{2\pi}{L}n_i, \quad {\rm for}\,\, i=x,y,z\,,
\end{equation}
with $n_i = 0, \pm 1, \pm 2$, etc. The spin and isospin degeneracies for each momentum state are $g_s=2$ for pure neutron matter (PNM) and $g_s=4$ for symmetric nuclear matter (SNM), respectively. Filling up the lowest Pauli-allowed momentum states leads to closed-shell configurations for $A=g_s \mathcal{N}$ nucleons with $\mathcal{N} = 1, 7, 19, 27, 33, 57, \ldots$. In the present work, we perform all calculations at the $\mathcal{N}= 66$ shell closure, which is empirically known to have small finite-size effects~\cite{Hagen:2013yba,Marino:2024tfp}.  Our IMSRG calculations are performed in a truncated single-particle model space defined by
\begin{equation}
n_x^2 + n_y^2 + n_z^2 \leq n_\text{max}^2\,.
\end{equation}
The size of the model space is then increased until the computed energy per nucleon is practically independent of $n_\text{max}$. For the largest model space considered, $n_\text{max}^2=34$, our results for the energy per nucleon of SNM are converged to the few-tens-of-keV level.

\subsection{Normal-ordered Hamiltonian}
\label{subsec:NOHam}
Starting from the Hamiltonian $\hat{H}=\hat{T} + \hat{V}^{(2)}+\hat{V}^{(3)}$, where $\hat{T}$ is the kinetic energy and $\hat{V}^{(2)}$ and $\hat{V}^{(3)}$ are NN and 3N interactions, we normal-order $\hat{H}$ with respect to the reference state $\refket$, which is taken to be the Slater determinant comprised of the lowest $A$ discrete momentum states,
\begin{equation}
\begin{aligned}\label{eq_noH}
    \hat{H}&=E+\hat{f}+\hat{\Gamma}+\hat{W} \\ &= E \begin{aligned}[t]&+ \sum_{pq} f_{pq}\{a^\dagger_p a_q\}\\&+ \frac{1}{4} \sum_{pqrs} \Gamma_{pqrs}\{a^\dagger_p a^\dagger_q a_s a_r\}
    \\&+\frac{1}{36}\sum_{pqrstu}W_{pqrstu}\{a^\dagger_p a^\dagger_q a^{\dagger}_r a_u a_t a_s\}\,,\end{aligned}
\end{aligned}
\end{equation}
where operators enclosed by curly braces obey $\langle\Phi|\{\cdots\}|\Phi\rangle = 0$, and the zero-, one-, two-, and three-body parts of the Hamiltonian are given by
\begin{align}\label{noe}
    E &= \sum_i T_{ii} + \frac{1}{2}\sum_{ij}V^{(2)}_{ijij}
    + \frac{1}{6}\sum_{ijk}V^{(3)}_{ijkijk}\\
    \label{nof}
    f_{pq} &= T_{pq} + \sum_{i}V^{(2)}_{piqi} +\frac{1}{2}\sum_{ij}V^{(3)}_{pijqij}\\
    \label{nogamma} \Gamma_{pqrs} &= V^{(2)}_{pqrs} +\sum_i V^{(3)}_{pqirsi} \\
    W_{pqrstu} &= V^{(3)}_{pqrstu}\,.
\end{align}
We use the convention that single particle indices $a,b,\ldots$ refer to unoccupied (particle) states in reference $\refket$, $i,j,\ldots$ refer to occupied (hole) states, and $p,q,\ldots$ run over the entire single particle basis. All of our calculations are carried out in the normal-ordered two-body (NO2B) approximation, which is obtained by setting $W_{qrstuv}=0$ in Eq.~\eqref{eq_noH}. We note that the NO2B truncation has been shown to be a good approximation for soft chiral EFT interactions, such as those used in the present work~\cite{Binder:2012mk,Roth:2011vt,Djarv:2021xjg}.

\subsection{IMSRG formalism}
\label{subsec:imsrgth}

The basic idea of the IMSRG is to construct a continuous unitary transformation $U(s)$ (where $U(0) \equiv \mathbf{1}$) and apply it to the Hamiltonian
\begin{equation}\label{eq:SRG}
H(s) = U^{\dagger}(s) H U(s) \equiv H_{d}(s) + H_{od}(s)\,,
\end{equation}
such that $H(s)$ is progressively driven to a block-diagonal form at large values of the flow parameter
\begin{equation}
\lim_{s\rightarrow \infty}H_{od}(s) = 0 \,.
\end{equation}
Here, $H_d$ and $H_{od}$ are the suitably defined diagonal and off-diagonal parts of the Hamiltonian. Differentiating Eq.~\eqref{eq:SRG} with respect to $s$ and defining the anti-hermitian generator $\hat{\eta}(s)\equiv \frac{dU(s)^{\dagger}}{ds}U(s)$, we obtain the operator flow equation
\begin{equation}\label{flow}
    \dv{s}\hat{H}(s) = [\hat{\eta}(s), \hat{H}(s)]\,.
\end{equation}

Note that if the initial $\hat{H}$ and $\hat{\eta}$ operators are truncated at the normal-ordered two-body level (NO2B) level, the commutator induces normal-ordered three-body (NO3B) terms, with higher-rank operators induced as one integrates to larger $s$-values. Therefore, to close the flow equations, we work in the IMSRG(2) approximation, which amounts to truncating all operators to the NO2B level.\footnote{%
For brevity, we use the term ``IMSRG'' to refer to our calculations in the IMSRG(2) approximation throughout this article.%
}

\begin{figure}[t]
    \centering
    \includegraphics[width=0.48\textwidth]{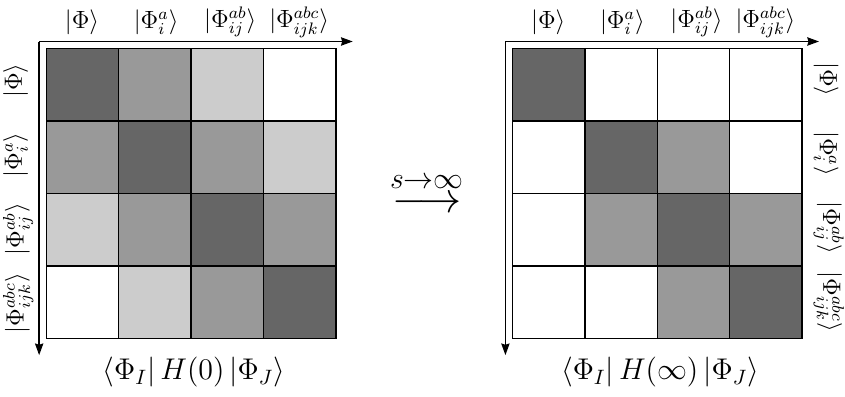}
    \caption{A schematic view of IMSRG decoupling \cite{Hergert:2016jk}.
    The many-body Hamiltonian is depicted in blocks, with the reference state $\refket$ and its particle-hole excitations (see the main text).
    During the IMSRG flow, the reference state is decoupled from its excitations.}
    \label{fig:decoupling}
\end{figure}

We wish to decouple $\refket$ from its particle-hole excitations as shown schematically in Fig.~\ref{fig:decoupling}, so that the flowing zero-body $E(s)=\langle\Phi|H(s)|\Phi\rangle$ becomes the ground state energy as $s\rightarrow\infty$. This is accomplished by adopting the White generator, which is given by\footnote{%
``\textrm{H.c.}'' stands for the Hermitian conjugate of the previous term.%
}

\begin{equation}
    \hat{\eta} = \frac{1}{4} \sum_{abij} \frac{\Gamma_{abij}}{\Delta_{abij}}\{a^\dagger_a a^\dagger_{b} a_{j} a_i\} -\textrm{H.c.},
\end{equation}
where
\begin{equation}
\begin{aligned}
    \Delta_{abij} &= \begin{aligned}[t]&\refbra \{a^\dagger_i a^\dagger_{j} a_{b} a_a\}\hat{H}\{a^\dagger_a a^\dagger_{b} a_{j} a_i\}\refket \\&- \refbra \hat{H} \refket\end{aligned}\\
    &= \begin{aligned}[t]f_{aa} &+ f_{bb} - f_{ii} - f_{jj} + \Gamma_{abab} + \Gamma_{ijij} \\ &- \Gamma_{aiai} - \Gamma_{bjbj} - \Gamma_{ajaj} - \Gamma_{bibi}\,.\end{aligned}
\end{aligned}
\end{equation}

Rather than direct integration of Eq.~\eqref{flow}, we use the Magnus expansion formulation~\cite{Morris:2015ve}, which offers a number of computational advantages. The idea is to express the IMSRG unitary transformation as a true exponential,
\begin{equation}
U(s) = e^{\Omega(s)}\,,
\end{equation}
where the anti-hermitian Magnus operator $\Omega(s)$ obeys the flow equation (suppressing explicit $s$ dependence for brevity)
\begin{equation}
    \dv{\Omega}{s}= \sum_{k=0}^{\infty} \frac{B_k}{k!}\mathrm{ad}^{(k)}_{\Omega}(\eta),
\end{equation}
where $B_k$ are the Bernoulli numbers, and the adjoint $\mathrm{ad}^{(k)}_{\Omega}(\eta)$ signifies a recursively defined  nested commutator:
\begin{equation}\label{eq:MagnusAdjointFlow}
    \mathrm{ad}^{(k)}_{\Omega}(\eta) = [\Omega,\mathrm{ad}^{(k-1)}_{\Omega}(\eta)]
    \quad\text{with}\quad
    \mathrm{ad}^{(0)}_{\Omega}(\eta) = \eta.
\end{equation}
The evolved Hamiltonian can then be expressed via the Baker--Campbell--Hausdorff formula
\begin{equation}\label{eq:MagnusBCH}
\begin{aligned}
    H(s) &= e^{\Omega(s)} H(0) e^{-\Omega(s)} \\
    & = \begin{aligned}[t]H(0) &+ [\Omega(s),H(0)] \\&+ \frac{1}{2}\left[\Omega(s),[\Omega(s),H(0)]\right] \\&+ \cdots\end{aligned}
    \end{aligned}
\end{equation}
In practice, the infinite series in Eqs.~\eqref{eq:MagnusAdjointFlow} and~\eqref{eq:MagnusBCH} are evaluated iteratively until the size of a given term falls below some numerical threshold.

\section{\pmmimsrg{} Emulation Framework}
\label{sec:method}

In this section, we present the formalism and implementation of the PMM emulator for IMSRG calculations of nuclear matter.
We employ a hybrid emulator composed of a rigorous implicit-reduced-basis method, a PMM, for the $s\rightarrow\infty$ Hamiltonian and a data-driven series expansion for the finite $s$ dependence of the IMSRG-calculated energies. The core principle guiding the emulator is a reduced-order model based on a reduced-basis method, with an unknown (implicit) basis informed by IMSRG calculations.

We consider the underlying free-space Hamiltonian up to N$^3$LO at fixed momentum cutoff, which is a function of the chosen chiral order indicated by ${\nu = \{0, 2, 3, 4\}}$ and associated low-energy constants (LECs) $\bc$,\footnote{%
The $c_i$'s are considered here generic LECs and are not to be confused with the $\pi$N LECs.%
}
which have the affine parameter dependence~\cite{Drischler:2022ipa}:
\begin{equation}
\label{eq:freespaceH}
H_\text{free}(\bc, \nu) = \sum_{j=0}^\nu\sum_{k=1}^{N_k} h_{jk}(\bc)V_k^{(j)}.
\end{equation}
That is, $H_\text{free}$ can be written as a sum of $N_k$ products of parameter-dependent functions $h_{jk}(\bc)$ and parameter-independent operators $V_k^{(j)}$, with $V_k^{(1)}\equiv 0$ in Weinberg power counting.

In this proof-of-principle work, we consider only the \ntwlo{} 3N LECs ${\bc \equiv [c_D, c_E, \dt, \ft]}$ and hold all other LECs constant at their best-fit values (i.e., $N_k = 4$ in Eq.~\eqref{eq:freespaceH}).
Consequently, the parameter dependence represented by the (smooth) functions $h_{jk}$ is linear; that is, $h_{jk}(c_i) = c_k$. We emphasize, however, that linearity is not a necessary requirement.
Following Ref.~\cite{Drischler:2017wtt}, we choose the order-by-order NN potentials developed by Entem, Machleidt, and Nosyk~\cite{Entem:2017gor} with the momentum cutoff $\Lambda=450~\MeV$ and combine them with 3N forces (if present) at the same chiral order.
Additionally, PNM does not depend on $c_D$ and $c_E$ for the nonlocal regulator functions considered~\cite{Hebeler:2009iv}. The in-medium Hamiltonian is then just a basis transformation parameterized by the number of allowed single particle states $N_s$, the proton fraction $Y_p$, and the nuclear density $n$. Denoting the Hamiltonian parameters by $X_H\equiv[\bc,\nu,N_s,Y_p,n]$, then
\begin{equation}
\label{eq:inmediumH}
    H_{\text{IM}}(X_H) = U^\dagger(N_s, Y_p, n) H_{\text{free}}(\bc, \nu) U(N_s, Y_p, n),
\end{equation}
where $U(N_s, Y_p, n)$ is a partial isometry---the operator which induces the projection $U^\dagger U$---parameterized by the aforementioned values.

In principle, the IMSRG result is the ground state energy of Eq.~\eqref{eq:inmediumH} plus corrections which are parameterized by all Hamiltonian parameters and which decay exponentially in the flow parameter, $s$,
\begin{equation}
\label{eq:flowed}
\begin{aligned}
E_0^\text{IMSRG}(X_H; s) =~ &E_0(X_H) \\ &+ \sum_{i=1}^r \alpha_i(X_H) \exp\left[-\beta_i(X_H) s\right],
\end{aligned}
\end{equation}
where $\alpha(X_H), \beta(X_H) \geq 0$ are smooth, non-negative functions parameterized by the Hamiltonian parameters, and $E_0(X_H)$ satisfies
\begin{equation}
\label{eq:evd}
H_\text{IM}(X_H)\ket{\Psi_0} = E_0(X_H)\ket{\Psi_0}.
\end{equation}

Consider now the subspace of a chosen dimension ${k_1\leq\text{dim}(H_\text{IM})}$ that best\footnote{\label{ftn:best}Here, ``best'' refers to optimality for the entire emulator's predictions. In this work, this refers only to the ground state energy.} preserves Eq.~\eqref{eq:evd}. Let $P_1$ be the partial isometry whose columns span this space. We can then compress Eq.~\eqref{eq:evd} into this space
\begin{equation}
\label{eq:evd_compressed}
h_\text{IM}(X_H)\ket{\psi_0} \approx E_0(X_H)\ket{\psi_0},
\end{equation}
where $h_\text{IM}(X_H) \equiv P_1^\dagger H_\text{IM}(X_H) P_1$ and $\ket{\psi_0} \equiv P_1^\dagger \ket{\Psi_0}$. Importantly, $E_0(X_H)$ is preserved under this transformation. We consider another subspace of a given dimension $k_2$ with ${k_1<k_2\leq\text{dim}(H_\text{free})}$ that best\footref{ftn:best} preserves Eq.~\eqref{eq:inmediumH}. As before, let $P_2$ be the partial isometry whose columns span this space. Compressing the $H_\text{free}(\bc, \nu)$ present in the expression for $h_\text{IM}(X_H)$ using $P_2$ yields
\begin{equation}
\label{eq:inmediumH_compressed}
h_\text{IM}(X_H)\approx u^\dagger(N_s, Y_p, n) h_\text{free}(\bc, \nu) u(N_s, Y_p, n),
\end{equation}
where $u(N_s, Y_p, n)\equiv P_2^\dagger U(N_s, Y_p, n) P_1$ and $h_\text{free}(\bc, \nu)\equiv P_2^\dagger H_\text{free}(\bc, \nu) P_2$.
It is straightforward to show that the functional form of $h_\text{free}(\bc, \nu)$, the compressed $H_\text{free}(\bc, \nu)$, is identical to the full-dimensional one in Eq.~\eqref{eq:freespaceH}, leading us to define ${v_i^{(j)}}$ of dimension $k_2$ just as before. The defining observation is that the preservation of $E_0(X_H)$ under these compressions is independent of the specific $P_1,\,P_2$. We can thus allow the bases of these reduced spaces to be \textit{implicit} by absorbing these partial isometries into our definitions of the reduced operators and learning these operators directly from data for $\left\{X_H, E_0(X_H)\right\}$ instead of explicitly performing the compressions. This is the fundamental concept behind PMMs and is what allows our emulator to retain as much information about the underlying physics as possible, even when knowledge of these subspaces is unavailable or prohibitively expensive.

Since the functional dependence of $U(N_s, Y_p, n)$ (and therefore $u(N_s, Y_p, n)$) is not known, we instead substitute a universal form able to approximate all possible forms, given a sufficiently large ``hidden model'' size. For such an operator which compresses from a space of dimension $N$ to one of dimension $k\leq N$,
\begin{equation}
\label{eq:universalU}
U_\text{univ}(x_i) \equiv \exp\left[i\sum^l_{j=1} f_j(x_i) M_j\right](:, 1\mathbin{:}k),
\end{equation}
where $M_j$ are Hermitian operators of dimension $N$, $f_j(x_i)$ are universal functions (e.g., the output of a neural network), and $A(:, 1\mathbin{:}k)$ denotes taking the first $k$ columns of $A$. For sufficiently large $l$ and expressive $f_j$, this can represent any parameterized partial isometry to arbitrary accuracy.

Finally, we model the flow-dependence of the IMSRG predictions via exponentially decaying corrections. The magnitude and decay rate of these corrections are learned in a purely data-driven way by fitting universal data-driven PMM forms~\cite{Cook2025}. Since the deployed emulator needs only to produce accurate predictions at $s\rightarrow\infty$, it is less important that the exact physics of these corrections is captured compared to that of $E_0(X_H)$.

\begin{figure}
\centering
\includegraphics[width=\linewidth]{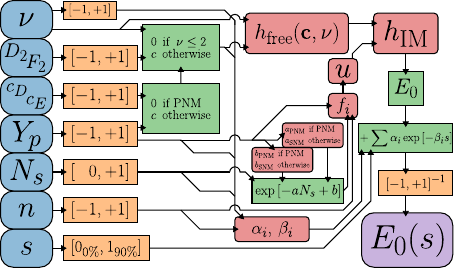}
\caption{\label{fig:pmm_diagram}Comprehensive diagram of the \pmmimsrg{} emulator. Inputs are shown on the left (blue boxes) and the predicted energy on the right (purple box). The rescaled chiral interaction and matter inputs feed into the construction of the free-space Hamiltonian $h_\text{free}$, while learned projections encode the dependence on the basis size, proton fraction, and density to form the reduced in-medium Hamiltonian $h_\text{IM}$. Red-shaded blocks indicate trainable components, including data-driven PMMs that capture exponential convergence with $N_s$ and finite-$s$ corrections. Diagonalization of $h_\text{IM}$ produces the $s\rightarrow\infty$ ground-state energy, which finite-$s$ corrections are added to and then rescaled to yield the emulator prediction $E_0(s)$.}
\end{figure}

The explicit process of using the PMM to make an emulated prediction is as follows. Given input parameters $c_D$, $c_E$, $\dt$, $\ft$, $\nu$, $N_s$, $Y_p$, $n$, and $s$, trained operators $v_i^{(j)}$, $m_1, m_2, \ldots, m_l \in \mathbb{H}(k_2)$, and trained scalar functions $f_1,\, f_2, \ldots, \,f_l, \alpha_1,\, \ldots, \alpha_r, \,\beta_1, \ldots, \beta_r$, first form $h_\text{free}(\bc, \nu)\in\mathbb{H}(k_2)$ using the form of Eq.~\eqref{eq:freespaceH}. Then, compute
\begin{equation}\label{eq:u_pmm}
    u(N_s, Y_p, n) = \exp\left[i\sum_{j=1}^l f_j(N_s, Y_p, n) m_j\right](:, 1\mathbin{:}k_1)
\end{equation}
and transform $h_\text{free}(\bc, \nu)$ as in Eq.~\eqref{eq:inmediumH_compressed} to form $h_\text{IM}(X_H)$. Perform an exact diagonalization of $h_\text{IM}(X_H)$ to find the emulated $E_0(X_H)$. The entire process can be done in $\bigO{k_2^3}$ time. If the emulated flow dependence is desired, one can evaluate all $\alpha_i$ and $\beta_i$ and then compute $E(X_H;s)$ as in Eq.~\eqref{eq:flowed}.

The emulation process, including data rescaling and preprocessing, is summarized in Fig.~\ref{fig:pmm_diagram}. In the diagram, model inputs ($\nu,\,\dt,\,\ft,\,c_D,\,c_E,\,Y_p,\,N_s,\,n,$ and $s$) are shown on the left. Components with trainable parameters are depicted with rounded corners and are shaded red. All inputs except $N_s$ and $s$ are scaled uniformly so that the training data lies within the range $[-1,+1]$. $N_s$ is scaled so that the training values span $[0, +1]$, and $s$ is scaled so that the minimum training value becomes $0$ and the $90^\mathrm{th}$ percentile becomes $1$. This rescaling is a standard step in machine learning that serves only to improve training speed and accuracy. The rescaled LECs are processed to zero out $c_D$ and $c_E$ for pure neutron matter and zero out all LECs when the chiral order is below \ntwlo{} to satisfy the physical constraints of the underlying interaction. The processed LECs and chiral order form $h_\text{free}$ following Eq.~\eqref{eq:freespaceH}. The number of allowed single-particle states, $N_s$, is scaled by a trainable exponential decay whose coefficients vary between PNM and SNM to help the model identify exponential convergence with $N_s$. Data-driven regression PMMs described in Ref.~\cite{Cook2025} identify hidden features $f_i$ for the basis projection $u$ from the rescaled $N_s$, $Y_p$, and $n$ as in Eq.~\eqref{eq:u_pmm}. The reduced-order in-medium Hamiltonian $h_\text{IM}$ is formed from $u$ and $h_\text{free}$ by Eq.~\eqref{eq:inmediumH_compressed} and diagonalized to find the ground state energy, $E_0$. All rescaled inputs except $s$ form hidden features $\alpha_i$ and $\beta_i$ from data-driven PMMs. These hidden features constitute the amplitude and decay scales of exponentially decaying finite-$s$ corrections to the ground-state energy. Finally, the rescaling of the training energies (which were scaled to be in $[-1,+1]$) is undone, yielding the emulator's prediction for $E_0(s)$.

The implementation, training, UQ, and deployment of our emulator are accomplished via the open-source \texttt{pyPMM} library~\cite{pyPMM2026}. We train the operators and universal functions of our emulator by minimizing a loss function using a modified implementation of the Adam gradient descent method for complex parameters~\cite{Adam}. Our loss function is an annealed average of the inverse-class-frequency-weighted mean squared error (wMSE) and the variance of the inverse-class-frequency-weighted error (wVAR). Let
\begin{equation}
\Delta\mleft(X_H;\,s\mright)\equiv E^\text{IMSRG}_0\mleft(X_H;\,s\mright) - E^\text{PMM}_0\mleft(X_H;\,s\mright)
\end{equation}
denote the prediction error. Then, the wMSE is given by
\begin{equation}
    \text{wMSE}\equiv\mean\limits_{\left(X_H;\,s\right)\in\mathcal{T}} w(X_H)\Delta\mleft(X_H;\,s\mright)^2
\end{equation}
where $\mathcal{T}$ is the training set\footnote{As is standard practice, we use the mini-batch variant of Adam. Thus, the training set is replaced by the mini-batch in all terms of the loss function during training. The weight $w(X_H)$ is unchanged.} and $w(X_H)$ is the inverse of the proportion of the training set which is represented by the specific $\nu$, $Y_p$, and $N_s$ values in $X_H$,
\begin{equation}
\hspace{-0.5em}w(X_H) \equiv \frac{\abs{\mathcal{T}}}{\abs{\left\{X_H^\prime\in\mathcal{T} \,\middle| \,\left(\nu,Y_p,N_s\right)=\left(\nu^\prime,Y_p^\prime,N_s^\prime\right)\right\}}}.
\end{equation}
This weighting ensures that data taken at a certain chiral order, proton fraction, or number of allowed single particle states do not dominate the model training simply due to over- or under-representation in the training set. Similarly, the wVAR is given by
\begin{equation}
\text{wVAR}\equiv\mean\limits_{\left(X_H;s\right)\in\mathcal{T}}w(X_H)\left[\Delta\mleft(X_H;\,s\mright) - \overline{\Delta}_w\right]^2,
\end{equation}
where $\overline{\Delta}_w$ is the weighted mean of the prediction error given by
\begin{equation}
\overline{\Delta}_w\equiv\frac{\sum\limits_{\left(X_H;s\right)\in\mathcal{T}} w(X_H) \Delta\mleft(X_H;\,s\mright)}{\sum\limits_{\left(X_H;s\right)\in\mathcal{T}} w(X_H)}.
\end{equation}
Note that only $\overline{\Delta}_w$ is normalized by the sum of the weights. This ensures that, during mini-batch gradient descent, particular mini-batches with larger---or smaller---than average weights do not disproportionally influence the emulator's parameters. The complete loss function used during training at Adam epoch $t$ is then
\begin{equation}
\frac{\xi\text{wMSE} + \gamma(t) \text{wVAR}}{\xi + \gamma(t)},
\end{equation}
where $\xi$ is a hyperparameter weighting the wMSE, and $\gamma(t)\in[0, 1]$ is the annealing given by
\begin{equation}
\gamma(t) = \clipfunc\mleft(\frac{t - t_\text{wMSE}}{t_\text{transition}}, 0, 1\mright),
\end{equation}
where $t_\text{wMSE}$ is a hyperparameter that determines how many epochs of pure wMSE loss ($\gamma(t) = 0$) to train the model with before gradually, at a rate determined by the hyperparameter $t_\text{transition}$, transitioning to a loss composed of $\xi$ parts wMSE and one part wVAR ($\gamma(t) = 1$). This choice of loss function allows training to quickly converge to an accurate but imprecise model before gradually encouraging an accurate and precise model. Training progress is quantified by the top $95^\text{th}$ percentile of the absolute error on the validation set, which we refer to as the validation loss.

The architecture, hyperparameters, and loss functions were chosen by standard partial grid search. Using the same entirely disjoint training and validation sets for each trial, models with different hyperparameters were trained via the aforementioned method and the model with the best validation loss across all trials was selected as our final emulator. A small amount of additional validation data was created afterward to manually verify the performance of the selected model. This additional data is not included in any of the sets described in this work.

Our implementation uses universal forms of PMMs for the $f_j,\alpha_i,$ and $\beta_i$ functions since the smoothness of the output can be tuned and model selection showed that such PMMs regularly outperformed similarly sized or larger feedforward neural networks. The trainable parameters of our emulator amount to fewer than $\num{30000}$ single-precision floating-point numbers\footnote{$h_\text{free}$ is a $17\times 17$ Hermitian matrix and $h_\text{IM}$ is a $7\times 7$ Hermitian matrix.} ($\SI{120}{\kilo\byte}$). With overhead, we are able to store our model and the training set which is required for the third term in the uncertainty heuristic---described in Section~\ref{sec:emu_uq}---in less than $\SI{1}{\mega\byte}$ and perform uncertainty-quantified inference with less than $\SI{2}{\giga\byte}$ of memory, making the trained model small enough and efficient enough to be deployed on virtually any hardware from the last decade.

\begin{figure}
\centering
\includegraphics[width=0.999\linewidth]{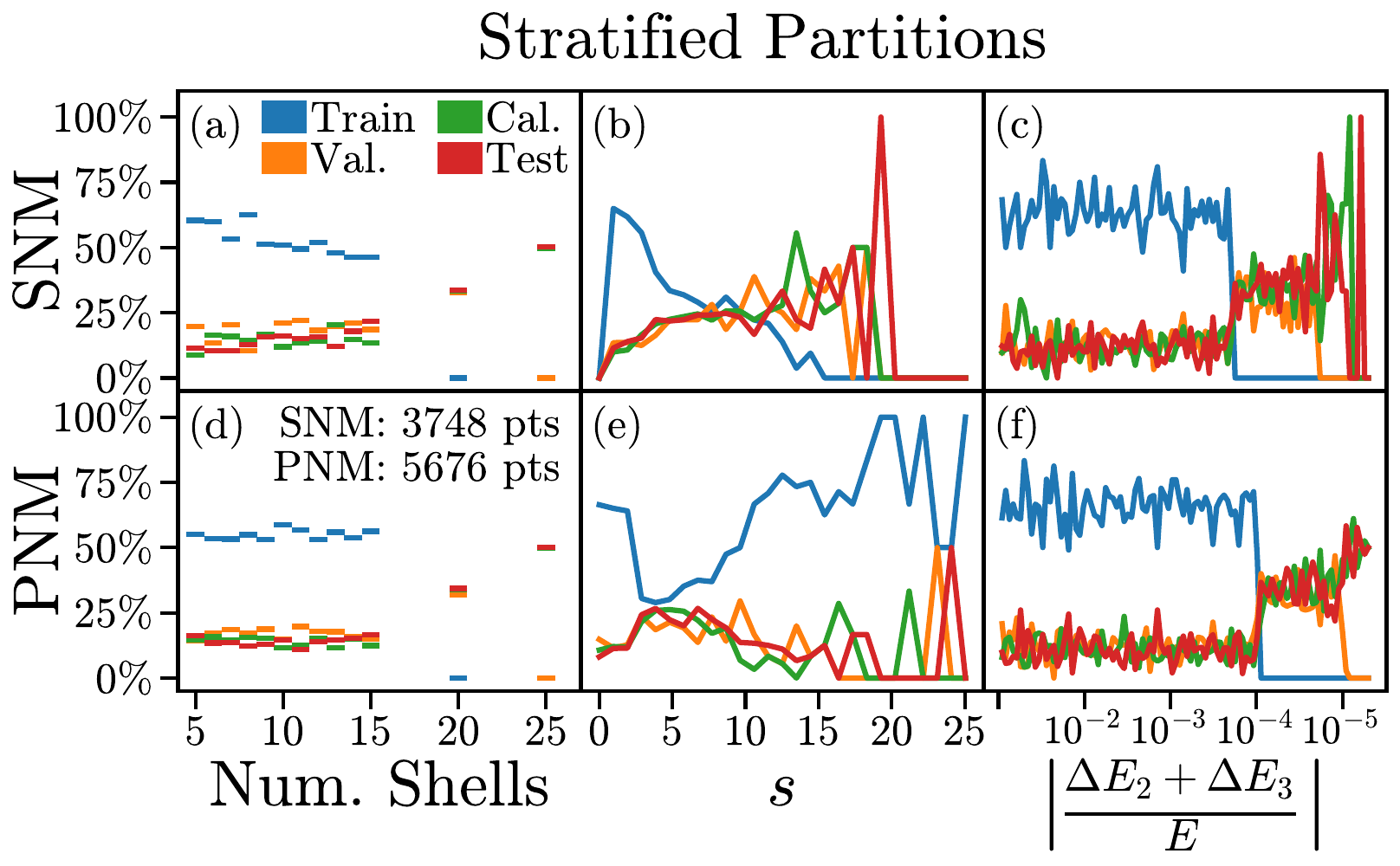}
\caption{\label{fig:partitions}The stratified partitioning of the IMSRG samples used for training, validating, calibrating, and testing our emulator. The training data (blue) is taken only from relatively unconverged samples with $15$ or fewer closed shells. Validation data (orange) are drawn from up to $20$ closed shells and include nearly converged samples. No fully converged samples or data from the largest number of closed shells, $25$, are present in the training or validation sets. The calibration (green) and testing (red) datasets span the full range of available IMSRG samples equally. All four datasets are entirely disjoint. Due to the comparative difficulty in obtaining IMSRG results for SNM, the number of SNM samples is roughly two-thirds that of PNM samples. The number of closed shells is determined by the number of single-particle states~\cite{Lietz2017}. $\Delta E_2$ and $\Delta E_3$ denote the second- and third-order MBPT energy corrections of the IMSRG-evolved Hamiltonian, respectively. The relative magnitude of their sum serves as a diagnostic of the convergence of the IMSRG calculation, with smaller values of $\left|(\Delta E_2+\Delta E_3)/E\right|$ indicating better convergence.}
\end{figure}

To accurately gauge real-world model performance, we evaluate the model on withheld test data taken disproportionately from more-converged (generally larger $s$) IMSRG results with larger $N_s$. Similarly, we encourage the training process to prefer models with better extrapolation performance by partitioning the available data so that the validation samples favor more-converged IMSRG results with larger $N_s$. We show the partitioning of the IMSRG data used to train, validate, calibrate, and test our emulator in Fig.~\ref{fig:partitions}.
Specifically, the top (bottom) row of Fig.~\ref{fig:partitions} shows the percentage of the data in each partition for SNM (PNM) calculations.
The three columns correspond, from left to right, to the number of closed shells, the flow parameter, and the perturbative convergence estimate of the IMSRG flow.
The latter estimate is obtained by performing second- and third-order many-body perturbation theory calculations using the IMSRG-transformed Hamiltonian $H(s)$, and assessing their perturbative contributions relative to the IMSRG energy, $E(s)=\langle\Phi|H(s)|\Phi\rangle$.
This convergence criterion is discussed in more detail in Ref.~\cite{Hergert:2016jk,Hergert:2017bc}.
At each point on the $x$-axis, the percentages across the four data partitions sum to 100\%.
For more details, see the caption of Fig.~\ref{fig:partitions}.

We can use our emulator to make predictions of the saturation density $n_\text{sat}$, saturation energy $E_\text{sat}$, symmetry energy $S$, slope parameter $L$, and incompressibility $K$. Denote $X_{\text{SNM or PNM}}\equiv\left[\bc,\,\nu,\,N_s,\,Y_e =0.5~\text{or}~0.0,\,n\right]$, then we solve
\begin{equation}
n_\text{sat} = \arginf\limits_{n\in\left(\min\limits_{n^\prime \in \mathcal{T}} n^\prime, \max\limits_{n^\prime \in \mathcal{T}} n^\prime\right)} E(X_\text{SNM}; s)
\end{equation}
for $n$ in the open interval bounded by the training dataset, $\mathcal{T}$. Denote $X_\text{SNM or PNM}^\text{sat}\equiv\left[\bc, \nu, N_s, 0.5~\text{or}~0.0, n_\text{sat}\right]$, then the emulator predictions for the saturation energy, symmetry energy, and its slope parameter in the standard quadratic approximation, and the incompressibility in SNM are respectively given by
{%
\newcommand{\sat}{\text{sat}}
\newcommand{\xpnm}{\ensuremath{X_\text{PNM}}}
\newcommand{\xsnm}{\ensuremath{X_\text{SNM}}}
\newcommand{\xsnmsat}{\ensuremath{\xsnm^\sat}}
\newcommand{\xpnmsat}{\ensuremath{\xpnm^\sat}}
\newcommand{\E}[1]{\ensuremath{E\mleft(#1\mright)}}
\newcommand{\evalat}[2]{\ensuremath{\left.#1\right|_{#2}}}
\begin{equation} \label{eq:sat_properties}
\begin{aligned}
    E_\text{sat} &= \E{\xsnmsat;s},\\
    S &= \E{\xpnmsat;s} - \E{\xsnmsat; s},\\
    L &= 3n_\sat\evalat{\frac{\partial}{\partial n}\Bigl[\E{\xpnm;s}-\E{\xsnm;s}\Bigr]}{n_\sat}\!,\\
    K &= 9 n_\sat^2 \evalat{\frac{\partial^2 \E{\xsnm;s}}{\partial n^2}}{n_\sat}.
\end{aligned}
\end{equation}%
}

\subsection{Emulator Uncertainty Estimation via Conformal Predictions}\label{sec:emu_uq}

\subsubsection{Conformal predictions}

We quantify our emulator uncertainties using split conformal predictions,\footnote{%
The proposed conformal-prediction-based estimation of emulator uncertainty is broadly applicable and not limited to IMSRG calculations of the nuclear EOS.%
}
which provide distribution-free finite-sample confidence guarantees that rely only on the exchangeability of data.\footnote{%
See Refs.~\cite{Dezdarani:2025aws} for recent applications of conformal predictions to quantifying uncertainties in nuclear scattering and the neutron star EOS.%
}
For a comprehensive introduction to this technique, we refer the reader to Refs.~\cite{Angelopoulos2023,Hulsman2022}. The core concepts are an uncertainty heuristic and a calibration dataset, which is disjoint from the training, validation, and testing datasets yet statistically identical to the testing dataset. Statistics of the emulator's performance on this calibration dataset are used to calibrate---or conformalize---the emulator's prediction set (confidence interval).

The uncertainty heuristic, $\unc{X}$, for a scalar-valued emulator is a scalar-valued function of the emulator's input data, which should be large when the emulator is uncertain and small when it is certain. The exact form and interpretation of the uncertainty heuristic can range from the standard deviation over an ensemble of emulators (the common ensemble learning or bootstrap aggregation method) to an additional model trained to estimate the residual. The key is that statistics of $\unc{X}$ over the calibration set can be used to conformalize the uncertainty heuristic via multiplicative corrections, producing half-interval widths for symmetric confidence intervals at any specified coverage probability (or, likewise, the miscoverage level). Denoting our emulator for the IMSRG calculations of the nuclear EOS by $E(X)$ (with $X\equiv\left(X_H; s\right)$) from data $\left\{\left(X_i;\hat{E}_i\right)\right\}$, the process is broadly as follows:
\begin{enumerate}
    \item Devise a function $\unc{X}$ which encodes a heuristic for the uncertainty of the emulator.
    \item Calculate the value of the score function---defined as the residual over the uncertainty heuristic---for each point in the calibration set
    \begin{equation}
        \mathcal{S}(X_i) = \frac{\left| E(X_i) - \hat{E}_i\right|}{\unc{X_i}}.
    \end{equation}
    \item For a desired miscoverage level $\alpha$ (or coverage probability $1-\alpha$), define $\hat{q}$ to be the
    \begin{equation}
        \frac{\left\lceil(1-\alpha)(n+1)\right\rceil}{n}
    \end{equation}
    quantile of the calibration set scores calculated in the previous step, where $n$ is the number of points in the calibration set and $\lceil\cdot\rceil$ is the ceiling function.
    \item $\hat{q}$ is the multiplicative correction to the uncertainty heuristic, and thus the conformalized model which returns the lower and upper bounds of the confidence interval with miscoverage level $\alpha$ for a test point $X$ denoted by $\mathcal{C}_\alpha(X)$ is
    \begin{equation}
        \mathcal{C}_\alpha(X) = \left\{
        \begin{aligned}
            E(X) &- \hat{q}\,\unc{X},\\ E(X) &+ \hat{q}\,\unc{X}
        \end{aligned}
        \right\}.
    \end{equation}
\end{enumerate}

Although the conformalized model has statically guaranteed coverage~\cite{Vovk1999}, the tightness---or usefulness---of the prediction intervals comes down to the ability of the uncertainty heuristic to accurately rank points in terms of expected uncertainty. We have deliberately included as much information about our emulator and the data it was trained on in our choice of uncertainty heuristic. The adaptive smoothness and analyticity of PMMs are particularly useful here in computing exact derivatives of our emulator, which inform the uncertainty heuristic.

Our uncertainty heuristic combines two sources of sensitivity and one source of uncertainty:
\begin{enumerate}
    \item The sensitivity of the trained emulator to perturbations in the trainable parameters,
    \begin{equation}
        S_\Theta\mleft(X\mright) = \frac{1}{\ell\mleft(\Theta\mright)}\left\Vert\left.\frac{\partial E}{\partial \Theta}\right|_{X} \odot \Theta \right\Vert_2^2,
    \end{equation}
    where $\odot$ is the elementwise product, $\Theta$ denotes the vector of all trainable parameters of the model, and $\ell\mleft(\Theta\mright)$ denotes the number of trainable parameters.
    \item The sensitivity of the trained emulator to perturbations in the input values,
    \begin{equation}
        S_{X}\mleft(X\mright) = \frac{1}{\ell\mleft(X_\text{cont}\mright)}\left\Vert\left.\frac{\partial E}{\partial X_\text{cont}}\right|_X \odot \delta\mleft(X_\text{cont}\mright)\right\Vert_2^2,
    \end{equation}
    where $X_\text{cont}$ denotes only the continuous input features, $\ell\mleft(X_\text{cont}\mright)$ denotes the number of continuous input features, and $\delta\mleft(X_\text{cont}\mright)$ denotes the characteristic scale of each of the continuous input features, which we take to be the inter-quartile range of each feature.
    \item The distance of the test point to the training set, $S_d(X)$, defined by the mean unweighted Gower's distance between the test point and all points in the training set within a distance threshold equal to the median unweighted Gower's distance between all pairs of points in the training set, excluding any features that can have infinite values ($s$ and $N_s$)~\cite{Gower1971}.
\end{enumerate}
The uncertainty heuristic for a test point is the normalized sum of these three contributions. Each term is normalized by its median absolute deviation over the calibration set.

\subsubsection{Accuracy of the emulator uncertainty estimate}

\begin{figure}
    \centering
    \includegraphics[width=\linewidth]{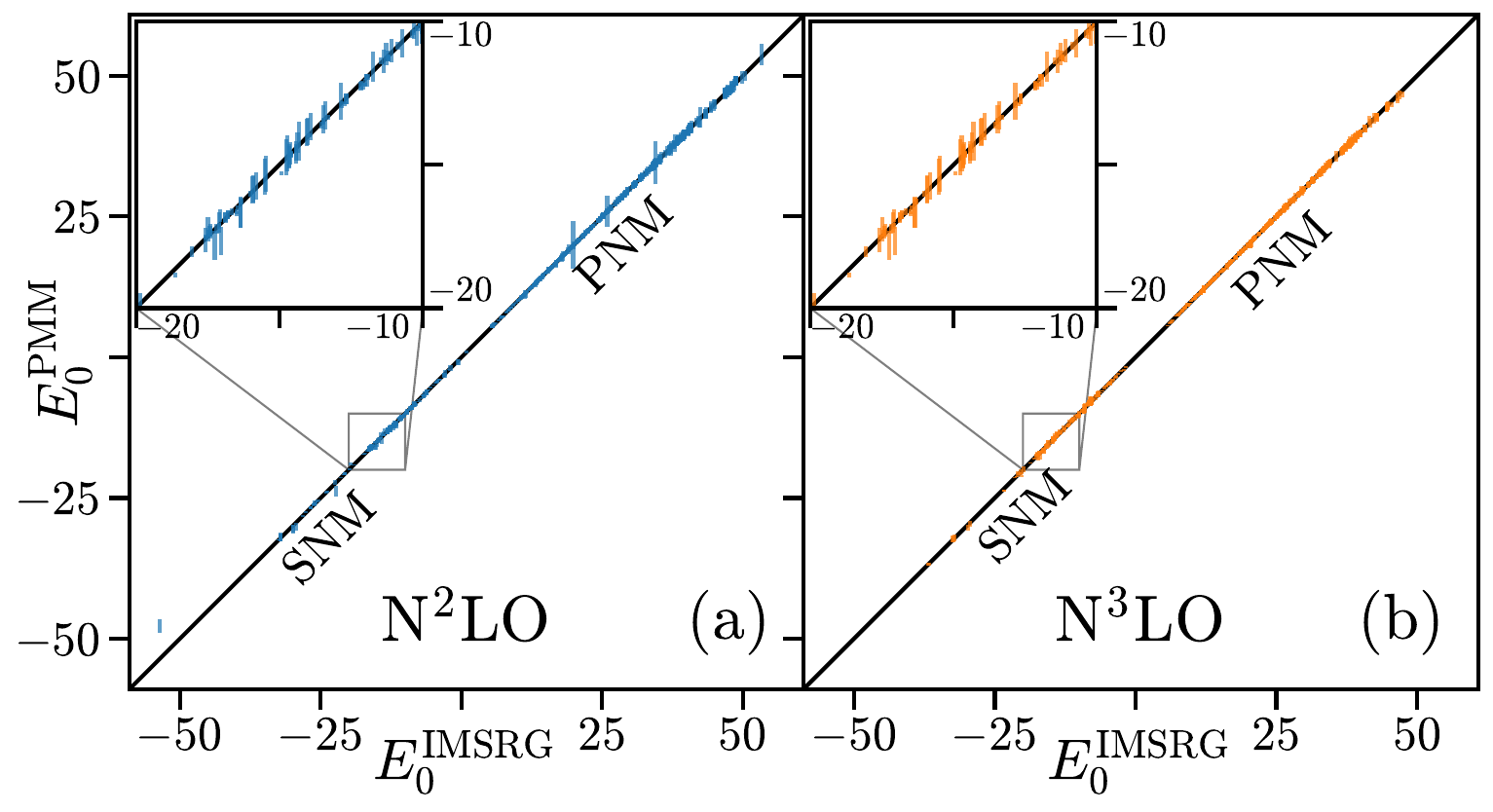}
    \caption{Comparison of PMM-predicted energies at the $95.45\%$ confidence level and withheld test IMSRG results for both PNM (here, $66$ neutrons) and SNM (here, $66$ protons and $66$ neutrons) as all degrees of freedom, including LECs and density, are varied for both \ntwlo{} (panel a) and \nthlo{} (panel b). Insets show additional detail around the empirical saturation energy.
    All values are in units of $\MeV$.}
    \label{fig:pmmaccuracy}
\end{figure}

We first demonstrate the accuracy of our emulator on withheld test data across both SNM and PNM at \ntwlo{} and \nthlo{} in Fig.~\ref{fig:pmmaccuracy}. The test data includes both SNM and PNM, varies all LECs considered in this work ($\cd$, $\ce$, $\dt$, $\ft$) and the nuclear density, and includes data of the most converged IMSRG results for the largest number of shells, neither of which were represented in the training or validation data as discussed in Fig.~\ref{fig:partitions}. We show the $95\%$ confidence interval for the predicted energy versus the IMSRG result, demonstrating both strong agreement and tight intervals.

\begin{figure}
\includegraphics[width=0.999\linewidth]{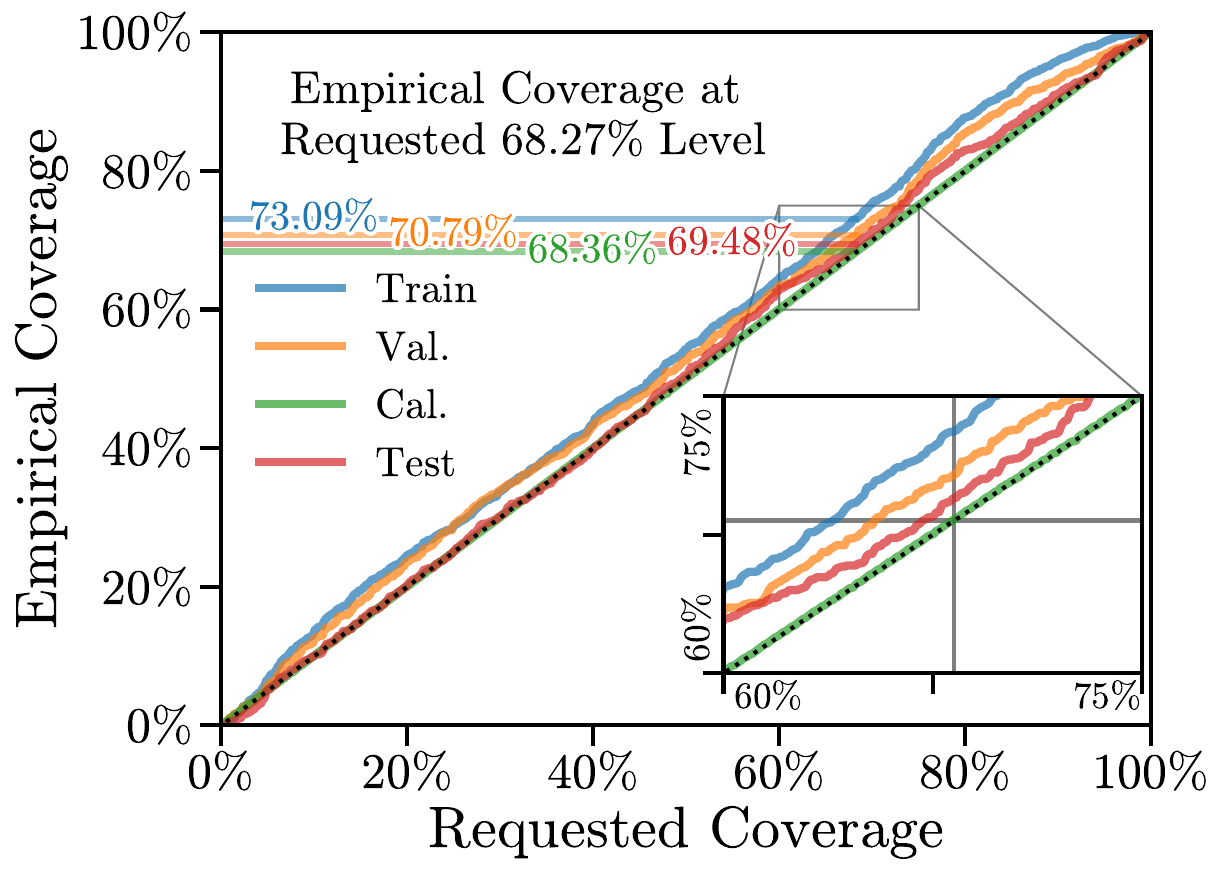}
\caption{\label{fig:model_coverage}Actual coverage versus requested confidence level for our emulator on each of the four data partitions. The observed coverages at the requested $68.27\%$ level are shown as horizontal lines, and the coverage range from 60 to 75\% is magnified in the inset to illustrate the small deviations from the ideal coverage observed.}
\end{figure}

We demonstrate the efficacy and trustworthiness of our UQ in Fig.~\ref{fig:model_coverage}, which shows the observed coverage versus the requested confidence level on the data. Following the split conformal prediction method, the empirical coverage on the calibration set is guaranteed to be exactly the requested coverage up to single-point over-coverage. If the uncertainty heuristic did not effectively rank points based on the true residual or the calibration dataset was not representative of unseen test data, then the empirical coverage on the withheld test data would deviate greatly from the requested coverage---yielding uninformative and untrustworthy uncertainty estimates. In contrast, Fig.~\ref{fig:model_coverage} shows that our UQ produces empirical coverages that track closely with the requested coverage and, for confidence levels greater than $10\%$, the maximum over-coverage is $3.45\%$ and the maximum under-coverage is $0.55\%$ on the withheld test data. While not guaranteed by split conformal predictions in practice, we observe that deviations from the requested coverage on the test data are most often toward over-coverage, yielding slightly overly cautious uncertainty estimates, which we find preferable to an overly confident model. Note that we separated the IMSRG data into training, validation, calibration, and test data sets, as is common in machine learning and indicated in the legend in Fig.~\ref{fig:model_coverage}.

When propagating emulator uncertainties to derived quantities, such as the incompressibility of SNM at saturation density, we allow for empirically verified smoothness assumptions about the emulator's energy predictions. This breaks the coverage guarantees while still producing highly accurate approximate intervals and greatly simplifying the uncertainty propagation.

We additionally use the calibration set for fixed, stratified bias correction. This does not change the model parameters, the distribution of residuals, or the widths of confidence intervals. Instead, we compute a constant offset that ensures that the median error on the calibration set is exactly $0$ for each category specified by the discrete values of $\nu$ and $Y_p$. While this correction is beneficial, we find that the largest-magnitude bias correction---which occurred for SNM at \nthlo{}---represented less than $0.037\%$ of the emulator's output range for the corresponding category.

\section{Results and discussion}
\label{sec:results}

In this section, we discuss our main results.
Following the methodology outlined in Sec.~\ref{sec:method}, we construct a \pmmimsrg{} emulator that is trained on IMSRG-generated PNM and SNM calculations encompassing a range of densities, basis sizes, and four chiral orders (i.e., up to \nthlo{}). Specifically, we perform IMSRG training calculations across $n = [0.08, 0.24] \, \fmiq$ and broad ranges $[-5, 5]$ for $c_D$ and $c_E$, $[-4.2, 4.2]~\mathrm{fm^4}$ for $D_2$ and $F_2$, based on the \nthlo{} Entem-Machleidt-Nosyk (EMN) NN interaction with a cutoff of 450~\MeV, combined with 3N forces at the same chiral order.

\subsection{Exploring saturation properties of 3N forces}
\label{subsec:general_sat_prop}

\begin{figure}
\centering
\includegraphics[width=0.999\linewidth]{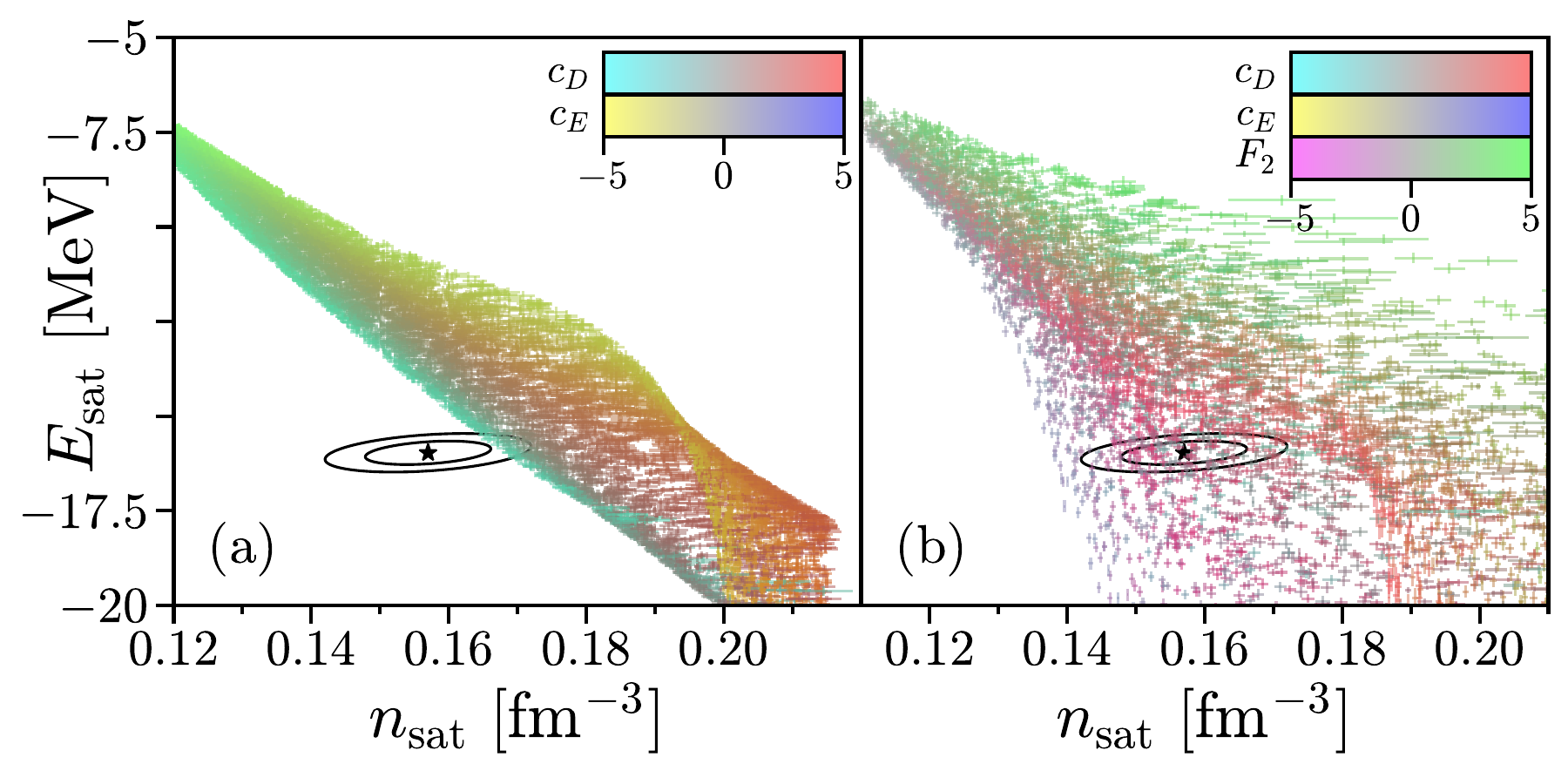}
\caption{\label{fig:coester} Saturation point distributions in SNM obtained from $\num{10000}$ uniform random samples in the range specified by the legends of $c_D$ and $c_E$ (panel a) and $c_D,\, c_E,$ and $F_2$ (panel b) with $\left\{N_s,\,s\right\}=\left\{\infty,\,\infty\right\}$ at \nthlo{}. The color of each point is determined by a color average of the corresponding colors given by the value of each LEC indicated in the legends. Each saturation point is shown with the associated confidence intervals at the $68.27\%$ level. The empirical saturation point~\cite{Drischler:2024ebw} is depicted by the star (mean value), and the associated $68.27\%$ (inner ellipse) and $95.45\%$ confidence-level contours (outer ellipse) are shown in black.
}
\end{figure}

Figure~\ref{fig:coester} shows the distributions of emulated saturation points ($n_\text{sat}, E_\text{sat}$) resulting from randomly sampling \num{20000} 3N LECs $\bc = \left[c_D,\, c_E \right]$ in panel~(a) and $\left[c_D,\, c_E,\, F_2\right]$ in panel~(b).
For simplicity, we sample all 3N LECs \emph{uniformly} in the broad range $[-5, +5]$, where we use $\fm^4$ as the units for $F_2$ (see the legends), while keeping the \nthlo{} EMN~450~\MeV\ NN potential and the \nthlo{} 3N forces constant.
The color of the samples shown in Fig.~\ref{fig:coester} is determined by the average color of the colors associated with the individual LECs given in the legends.
The empirical saturation point, which is given by a bivariate student-$t$ distribution and depicted by the star (i.e., mean value) and two confidence ellipses at $68\%$ and $95\%$ credibility, respectively, was obtained in Ref.~\cite{Drischler:2024ebw} by Bayesian model mixing of the empirical constraints of the saturation points obtained from various Skyrme and relativistic mean field functionals.
As shown in panel~(a) of Fig.~\ref{fig:coester}, this sampling strategy results in a Coester-like anti-correlation band~\cite{Coester:1970ai}.
However, with only $c_D$ and $c_E$ as degrees of freedom, this Coester band is markedly shifted to higher energies per particle and densities relative to the empirical points, consistent with the findings in the literature (e.g., see Ref.~\cite{Drischler:2021kxf}).

Motivated by this well-known shortcoming, we add a quark-mass-dependent 3N force, which was recently identified to appear at \ntwlo{}, $F_2$.
Previous studies~\cite{Cirigliano2025Chiral,Vernik:2025czp} indicate that the contribution from $F_2$ is more significant than that from $D_2$, a trend that we have verified with our IMSRG calculations.
Accordingly, we follow Ref.~\cite{Vernik:2025czp} and focus on $F_2$ throughout the remainder of this work; i.e., $D_2=0$.
As shown in panel~(b) of Fig.~\ref{fig:coester}, we find that including $F_2$ enables, while significantly widening the Coester band, these forces to saturate closer to the empirical point.
This improved agreement motivates us to employ the \pmmimsrg{} emulator to investigate more systematically the LEC dependence of the EOS on $F_2$ and to further constrain its value in Sec.~\ref{subsec:constrain}.

\subsection{Variations of $c_D$ and $c_E$ about the triton binding energy}

Next, we explore parametric uncertainties in the NN and 3N interactions constructed in Ref.~\cite{Drischler:2017wtt} by fitting $c_D$ and $c_E$ to the triton binding energy and the empirical saturation point using MBPT calculations.
To this end, we consider the $c_E(c_D)$ correlation due to the triton binding energy, which is given in Figure~1 in the supplemental material of Ref.~\cite{Drischler:2017wtt}, and assume a somewhat arbitrary symmetric uncertainty about the blue-dashed line in that Figure~1, as described below.
In future work, this additional uncertainty could be due to EFT truncation errors in the triton binding energy predictions, which were not considered in Ref.~\cite{Drischler:2017wtt}.
We construct a Gaussian process (GP)~\cite{rasmussen2006gaussian} with a fixed mean function and covariance function (or kernel) to represent this uncertainty band.
Specifically, we use a cubic spline interpolating the discrete $c_D(c_E)$ data obtained in Ref.~\cite{Drischler:2017wtt} as the mean function and choose constant values for the hyperparameters of the radial basis function (RBF) kernel:
\begin{equation}
    \kappa_\text{RBF}(c_D,c_D') = \sigma_f^2 \exp \left[- \frac{(c_D-c_D')^2}{\ell^2} \right],
\end{equation}
with variance $\sigma_f^2 = 0.1^2$ and correlation length $\ell = 10$.

\begin{figure}[tb]
    \centering
    \includegraphics[width=\linewidth]{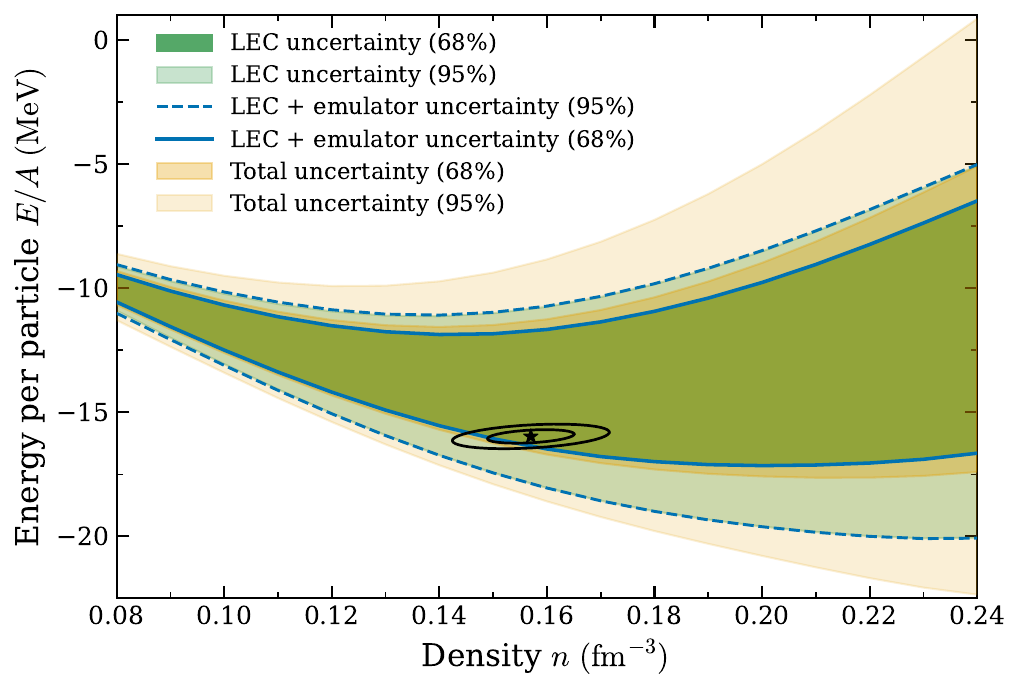}
    \caption{%
    Confidence regions of the EOS in the limit of SNM at \nthlo{} when varying the widened $c_D$-$c_E$ correlation that fits the triton binding energy as obtained in Ref.~\cite[see Fig.~1 in the Supplemental Material]{Drischler:2017wtt}.
    The error bands are obtained through ordinary Monte Carlo sampling from the probability distributions of the emulator and EFT truncation errors.
    See the main text for details.%
    }
    \label{fig:snm_cd_ce_variationmc}
\end{figure}

\begin{figure}[tb]
    \centering
    \includegraphics[width=\linewidth]{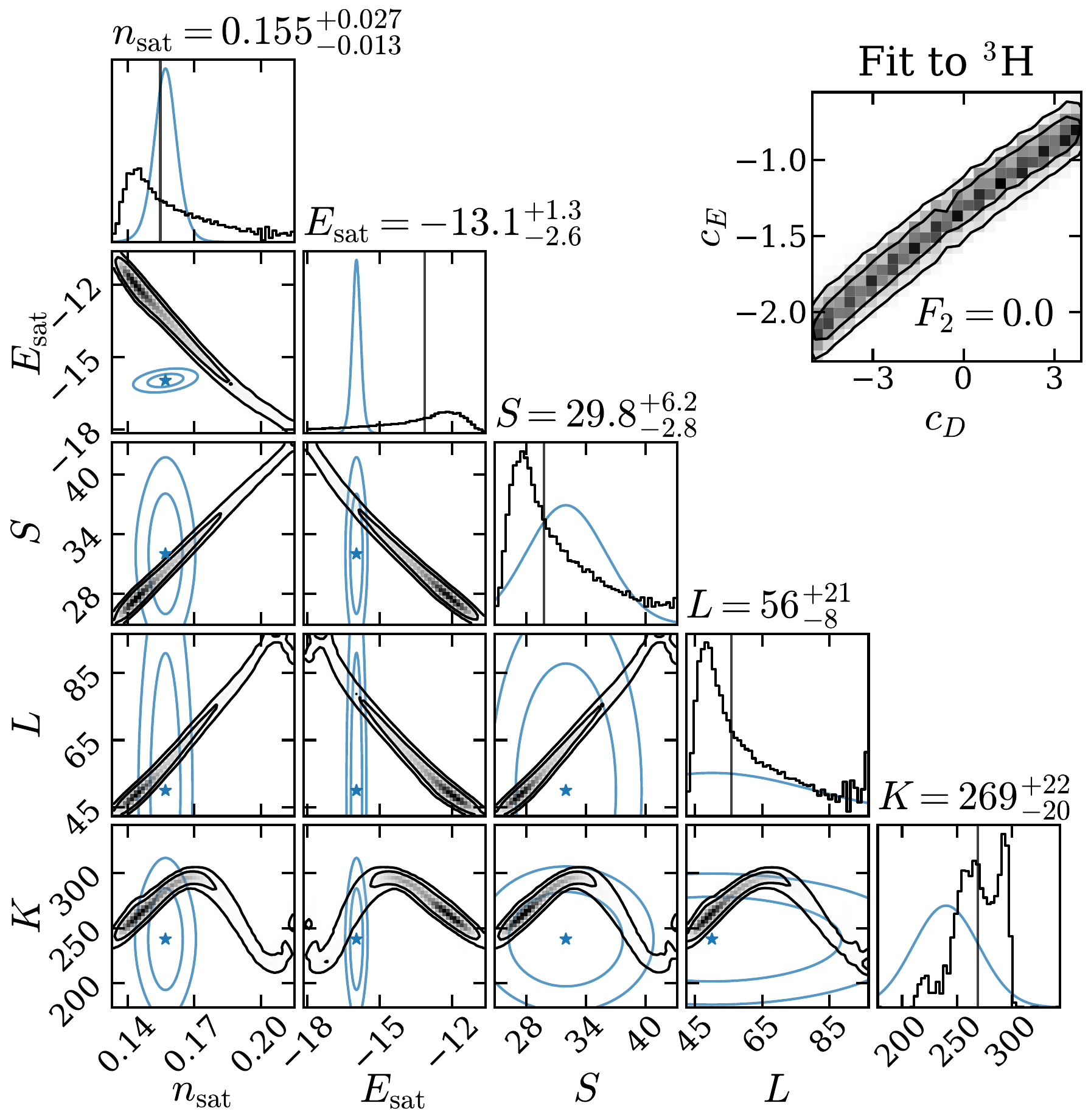}
    \caption{%
    Low-density EOS parameters associated with the sampled EOSs depicted in Fig.~\ref{fig:snm_cd_ce_variationmc} (black).
    The units are MeV, except for $n_\text{sat}$, which is in \fmiq.
    The blue lines and stars depict the empirical ranges defined in Eq.~\eqref{eq:emp_params}.
    The reported uncertainties in the titles correspond to $68\%$ credibility intervals centered on the median. %
    }
    \label{fig:snm_cd_ce_variation_eos_params}
\end{figure}

We evaluate the GP at \num{100} equally spaced $c_D$ values over the interval $[-5,5]$ and generate \num{100} $c_E(c_D)$ sample curves. This procedure yields a total of \num{10000} $(c_D,c_E)$ pairs.
Figures~\ref{fig:snm_cd_ce_variationmc} and~\ref{fig:snm_cd_ce_variation_eos_params} show the resulting SNM and associated low-density EOS parameters.\footnote{%
We do not show the PNM EOS because its energy per particle is independent of the 3N LECs considered~\cite{Hebeler:2009iv}, so varying these LECs would yield a single line rather than an uncertainty band.%
}

In Figure~\ref{fig:snm_cd_ce_variationmc}, we use the \pmmimsrg{} emulator to predict the EOS at all four chiral orders for each of the \num{10000} sampled \((c_D,c_E)\) pairs. We construct an approximate quantile function for the emulator error at each chiral order, using the conformal intervals over a range of miscoverage levels.
Assuming the emulator errors are independent across chiral orders, we draw samples from the emulator error distributions using inverse-transform sampling.
For each \((c_D,c_E)\) pair, we draw \num{100} emulator error samples at each chiral order and add them to the corresponding emulator predictions.
For each resulting set of emulator-error-adjusted EOS predictions, we use the BUQEYE pipeline~\cite{Melendez:2017phj,Melendez:2019izc}, described in more detail in section~\ref{subsec:uncertainty}, to determine the \nthlo{} EFT truncation error distribution at each density point and draw an additional \num{100} EFT truncation error samples. This ordinary Monte Carlo procedure therefore produces \(10^4\times10^2\times10^2=10^8\) EOS samples, incorporating uncertainties from the LECs, emulator errors, and EFT truncation errors. The LEC uncertainty bands (depicted in green) are obtained by taking pointwise quantiles over the EOS distribution associated with the sampled LECs after marginalizing over the emulator and EFT truncation errors. The combined LEC and emulator uncertainty bands (depicted in blue) are obtained by taking pointwise quantiles after marginalizing over the EFT truncation error. Finally, the total uncertainty bands (depicted in orange) in Fig.~\ref{fig:snm_cd_ce_variationmc} are obtained from the pointwise quantiles of the full EOS sample, including the LEC, emulator, and EFT truncation uncertainties.

We find that, even when relaxing the requirement to exactly reproduce the triton binding as discussed, these chiral NN and 3N forces still cannot reproduce the empirical saturation point at the $68\%$-level or better.
However, as shown in the correlation plot in Fig.~\ref{fig:snm_cd_ce_variation_eos_params}, the predicted Coester band is slightly shifted toward the empirical saturation point, while the other low-density EOS parameters explore a wide range of values.
For example, the symmetry energy $S = 29.8_{-2.8}^{+6.2}\, \MeV$ and its slope parameter $L = 56_{-8}^{+21}\, \MeV$ at $n_\text{sat}$.
In the following, we study the consequence of adding the $F_2$ term to the \ntwlo{} 3N forces.
To this end, we choose to constrain its coupling constant to empirical saturation properties.

\subsection{Calibrating 3N forces to empirical saturation properties with emulator uncertainties}
\label{subsec:constrain}

We devise a Bayesian parameter-estimation framework based on the \pmmimsrg{} emulator, with emulator uncertainties quantified using conformal prediction. In this exploratory study, we account only for emulator uncertainty and defer the inclusion of EFT truncation errors to future work.\footnote{%
LEC uncertainties and the EFT truncation errors are correlated. While neglecting these correlations may lead to misestimation of EFT predictions, rigorously quantifying and incorporating these uncertainties remains a work in progress~\cite{Carter:2026azo}.%
}
For a given set of 3N LECs $\bc = \left[c_D,\, c_E,\, F_2\right]$, the emulator efficiently generates the EOS in the limits of PNM and SNM.
The saturation properties are obtained by locating the minimum of the SNM EOS and evaluating the relevant derivatives at that point (see Eqs.~\eqref{eq:sat_properties} for the expressions), with associated emulator uncertainties propagated accordingly.

\subsubsection{Bayesian fit protocol}

The empirical observables used to constrain these LECs include the saturation density ($n_{\mathrm{sat}}$), saturation energy ($E_{\mathrm{sat}}$), symmetry energy ($S$), slope of the symmetry energy ($L$), and incompressibility of SNM ($K$). For the first two parameters, we adopt the joint bivariate student-$t$ distribution, $t_{9}(\boldsymbol{\mu}, \Psi)$, obtained in Ref.~\cite{Drischler:2024ebw}, with 9 degrees of freedom and mean value and scale matrix, respectively, given by (in units of \fmiq\ and \MeV)\footnote{
This distribution function for the empirical saturation point is depicted throughout this article in figures showing the SNM EOS.%
}
\begin{equation}
\boldsymbol{\mu} \approx
\begin{bmatrix}
0.157 \\
-15.97
\end{bmatrix},
\qquad
\Psi \approx
\begin{bmatrix}
0.005^2 & 0.017^2 \\
0.017^2 & 0.17^2
\end{bmatrix}.
\end{equation}
At the prior level, we assume that $S$, $L$, and $K$ have the independent normal distributions:
\begin{equation}\label{eq:emp_params}
\begin{aligned}
S &\sim \mathcal{N}(32,\, 3^2), \\
L &\sim \mathcal{N}(60,\, 25^2), \\
K &\sim \mathcal{N}(240,\, 20^2),
\end{aligned}
\end{equation}
where the mean values and standard deviations are given in MeV.
The priors for $S$ and $L$ are chosen to encompass, at the prior's $\approx 2\sigma$ level~\cite{Drischler:2026vdm}, the relatively large mean values from the recent PREX--II-informed extraction~\cite{Reed:2021nqk}, while the prior for $K$ is chosen to be consistent to (but slightly less informed than) the one obtained in Ref.~\cite{Roca-Maza:2018ujj} based on several analyses of isoscalar giant monopole resonances.
We adopt uniform priors for the dimensionless $c_D$ and $c_E$ over the range $[-5,\,5]$ to impose naturalness, and for $F_2$ over the range $[-(5f_{\pi}^4)^{-1},\,(5f_{\pi}^4)^{-1}]$,\footnote{%
This range corresponds to $|F_2| \lesssim  4.2 \,\mathrm{fm}^{4}$, so slightly smaller than the range used to generate the results for panel~(b) in Fig.~\ref{fig:coester}.%
}
with the pion decay constant $f_{\pi}$, as determined by renormalization-group analysis~\cite{Cirigliano2025Chiral}.

We consider the statistical discrepancy model of the emulator error:
\begin{equation} \label{eq:discrepancy_model}
    y_{\mathrm{IMSRG}} = y_{\mathrm{emu}} + \delta y_{\mathrm{emu}},
\end{equation}
where $y_{\mathrm{emu}}$ denotes the emulator's approximation to the high-fidelity prediction $y_{\mathrm{IMSRG}}$.
The difference between the two is given by the emulator error $\delta y_{\mathrm{emu}}$.
We emphasize that $y_{\mathrm{IMSRG}}$ is not the ``true'' value of the observable in nature, but rather the high-fidelity IMSRG prediction that we seek to emulate; it therefore carries additional theoretical uncertainties, e.g., from many-body approximations and EFT truncations, which we omit in the following.
Since $\delta y_{\mathrm{emu}}$ is not directly observed in practice, we model it as a random variable.

With the discrepancy model~\eqref{eq:discrepancy_model}, the likelihood of observing saturation properties $y_{\mathrm{IMSRG}}$ for a given set of LECs $\bc$ is obtained by marginalizing the emulator error via the convolution integral~\cite{Giri2026Emulator}
\begin{equation}
\label{eq:likelihoodconvolution}
\begin{aligned}
p(y_{\mathrm{IMSRG}} \mid \bc)
=
\int
&p\left(
y_{\mathrm{IMSRG}}-\delta y_{\mathrm{emu}}
\right)\\
&\times p(\delta y_{\mathrm{emu}})
\dd(\delta y_{\mathrm{emu}}),
\end{aligned}
\end{equation}
where $p(\cdot)$ indicates probability density distributions.
The likelihood~\eqref{eq:likelihoodconvolution} can be rewritten in terms of the quantile function of the emulator error, $q(u)$ with $u \in [0,1],$\footnote{%
The quantile function \(q(u)\) depends on the input LECs \(\bc\). For notational simplicity, we write \(q(u)\) in place of \(q_{\bc}(u)\) throughout this paper.
} %
using the probability integral transform~\cite{casella2002statistical}\footnote{%
See Theorem~2.1.10 in Ref.~\cite{casella2002statistical} for more details.%
}%
\begin{equation} \label{eq:likelihoodconvolution_trafo}
 p(y_{\mathrm{IMSRG}} \mid \bc)  =
\int_0^1
p\left(
y_{\mathrm{IMSRG}} - q(u)
\right)
\, \dd u.
\end{equation}
Here, we use a set of conformal intervals over a range of $\alpha$ values to approximate the quantile function $q(u)$ in Eq.~\eqref{eq:likelihoodconvolution_trafo}, as follows.
For a given miscoverage level $\alpha \in (0,1)$, conformal prediction yields an interval $(a_\alpha, b_\alpha)$ that satisfies
\begin{equation}
\begin{aligned}
p\big(a_\alpha \le y_{\mathrm{IMSRG}} \le b_\alpha\big) \approx 1 - \alpha.
\end{aligned}
\end{equation}
The confidence intervals are central by construction, implying that the CDF for $F_{y}$ satisfies
\begin{equation}
\begin{aligned}
F_{y}(a_\alpha) \approx \frac{\alpha}{2},
\quad
F_{y}(b_\alpha) \approx 1 - \frac{\alpha}{2},
\end{aligned}
\end{equation}
which, together with Eq.~\eqref{eq:discrepancy_model}, results in the shifted CDF for $\delta y_\mathrm{emu}$ evaluated at the end points:
\begin{equation}
\begin{aligned}
F_{\delta}(a_\alpha - y_{\mathrm{emu}}) \approx \frac{\alpha}{2},
\quad
F_{\delta}(b_\alpha - y_{\mathrm{emu}}) \approx 1 - \frac{\alpha}{2}.
\end{aligned}
\end{equation}
The corresponding quantile function is then given by $q(u) = F_{\delta}^{-1}(u)$, with $u \in (0,1)$, and thus
\begin{equation}
\begin{aligned}
q\left(\frac{\alpha}{2}\right) \approx a_\alpha - y_{\mathrm{emu}},
\quad
q\left(1 - \frac{\alpha}{2}\right) \approx b_\alpha - y_{\mathrm{emu}}.
\end{aligned}
\end{equation}
Computing the conformal intervals $[a_\alpha, b_\alpha]$ over a range of $\alpha$ values allows us to estimate the corresponding values of $q(u)$ on a grid of $u$ and then numerically interpolate them to obtain $q(u)$ over the interval $(0,1)$ for each given set of LECs $\bc$.
Specifically, we use an equidistant grid of $N_\alpha =500$ values of $\alpha \in (0,1)$, resulting in $N_\alpha$ central confidence intervals, from which we obtain the quantile function evaluated at $\alpha/2$ and $1-\alpha/2$ simultaneously.
Hence, we obtain $q(u)$ sampled at $2 N_\alpha = 1000$ grid points.
We then use linear interpolation between these points to approximate the full quantile function while preserving its monotone non-decreasing structure.

Let \(X(\bc)\) denote the central emulator prediction of the saturation
property \(X\) for a given set of LECs \(\bc\), and let \(q_X(u)\) denote
the corresponding quantile function of the emulator error, where
\[
X\in
\left\{
n_{\mathrm{sat}},
E_{\mathrm{sat}},
S,
L,
K
\right\}.
\]
We collect the empirical saturation point into the vector
\begin{equation}
\mathbf{y}_{nE}^{\mathrm{emp}}
\equiv
\begin{pmatrix}
n_{\mathrm{sat}}^{\mathrm{emp}} \\
E_{\mathrm{sat}}^{\mathrm{emp}}
\end{pmatrix},
\end{equation}
the central emulator predictions of \(n_{\mathrm{sat}}\) and
\(E_{\mathrm{sat}}\) into
\begin{equation}
\boldsymbol{\mu}_{nE}(\bc)
\equiv
\begin{pmatrix}
n_{\mathrm{sat}}(\bc) \\
E_{\mathrm{sat}}(\bc)
\end{pmatrix},
\end{equation}
and their emulator-error quantile functions into
\begin{equation}
\mathbf{q}_{nE}(u)
\equiv
\begin{pmatrix}
q_n(u) \\
q_E(u)
\end{pmatrix}.
\end{equation}
For a given \(u\in(0,1)\), we define the emulator-error-adjusted empirical saturation point by
\begin{equation}
\Delta\mathbf{y}_{nE}(u)
\equiv
\mathbf{y}_{nE}^{\mathrm{emp}}
-
\mathbf{q}_{nE}(u).
\end{equation}
For the remaining saturation properties, we define
\begin{equation}
\Delta X(u)
\equiv
X^{\mathrm{inf}}-q_X(u),
\qquad
X\in\{S,L,K\}.
\end{equation}
Here, \(X^{\mathrm{inf}}\) denotes the adopted prior-level calibration mean for the saturation property \(X\).
The complete calibration-data vector is
\begin{equation}
\mathbf{y}_{\mathrm{cal}}
\equiv
\left(
n_{\mathrm{sat}}^{\mathrm{emp}},
E_{\mathrm{sat}}^{\mathrm{emp}},
S^{\mathrm{inf}},
L^{\mathrm{inf}},
K^{\mathrm{inf}}
\right)^{\intercal}.
\end{equation}
Assuming that, conditional on \(\bc\) and \(u\), \((n_{\mathrm{sat}},E_{\mathrm{sat}})\) is independent of \((S,L,K)\), and that \(S\), \(L\), and \(K\) are mutually
independent, the likelihood of the calibration data, marginalized over the
emulator uncertainty, is
\begin{equation}
\label{eq:posterior_ycal}
\begin{aligned}
p\mleft(
\mathbf{y}_{\mathrm{cal}}
\;\middle|\;
\bc
\mright)
=
\int_0^1
&
t_9\mleft(
\Delta\mathbf{y}_{nE}(u)
\;\middle|\;
\boldsymbol{\mu}_{nE}(\bc),
\Psi
\mright)
\\
&\times
\mathcal{N}^{(S,L,K)}(u)\,\dd u,
\end{aligned}
\end{equation}
where
\begin{equation}
\mathcal{N}^{(S,L,K)}(u)\equiv \prod_{X\in\{S,L,K\}} \mathcal{N}\mleft( \Delta X(u) \;\middle|\; X(\bc), \sigma_X^2 \mright).
\end{equation}

\begin{figure}
    \centering
    \includegraphics[width=0.999\linewidth]{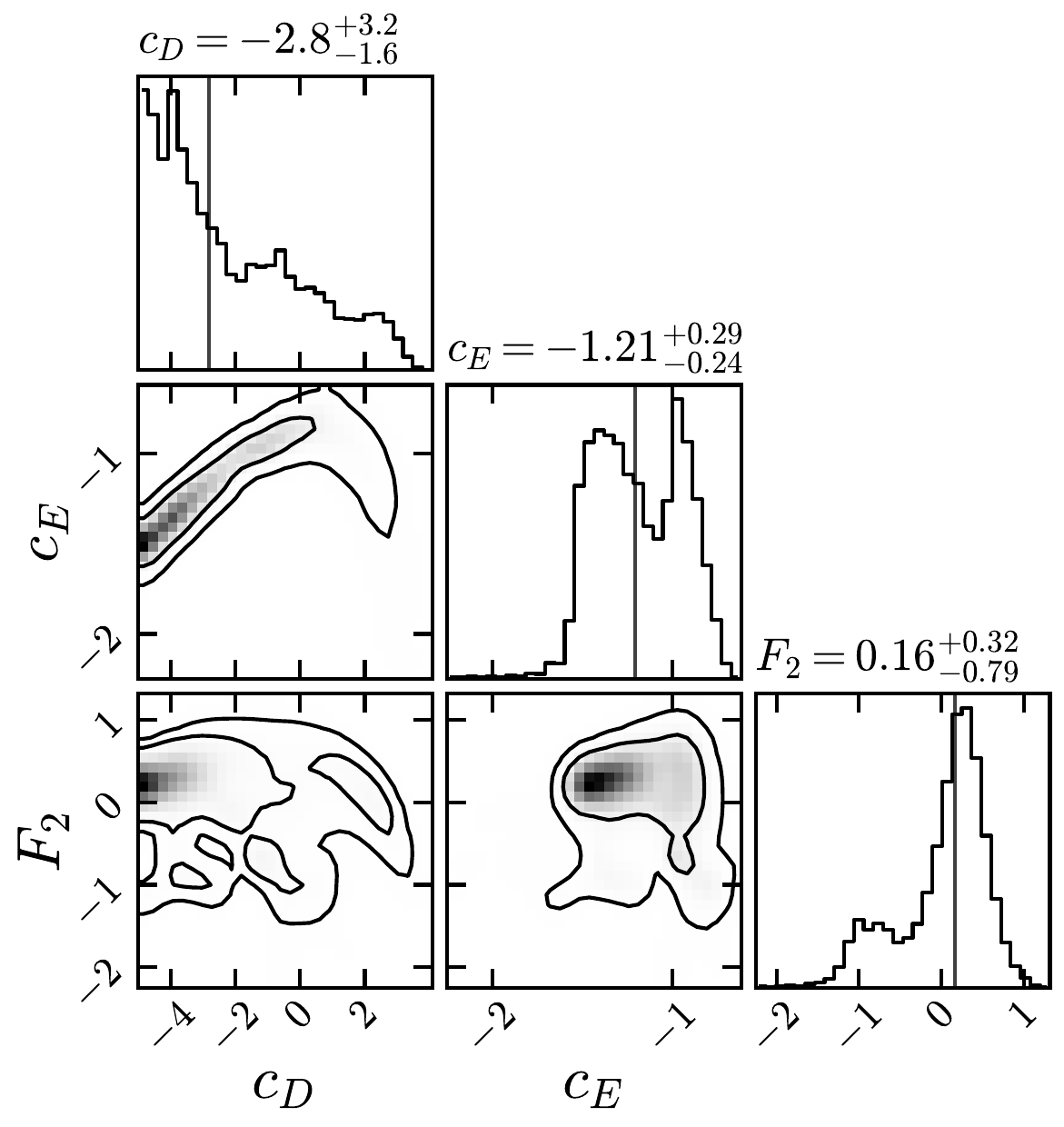}
    \caption{Corner plot showing the posterior distributions and pairwise correlations of the low-energy constants $c_D$, $c_E$, and $F_2$ obtained from the \pmmimsrg{} Bayesian analysis. The LECs $c_D$ and $c_E$ are dimensionless, whereas $F_2$ is given in $\mathrm{fm}^{4}$. The reported uncertainties in the titles correspond to 68\% credibility intervals centered on the median.
    Note that the uniform prior enforces the constraint $|c_D \leq 5|$.
    }
    \label{fig:cornerlec}
\end{figure}

\begin{figure}
    \centering
    \includegraphics[width=\linewidth]{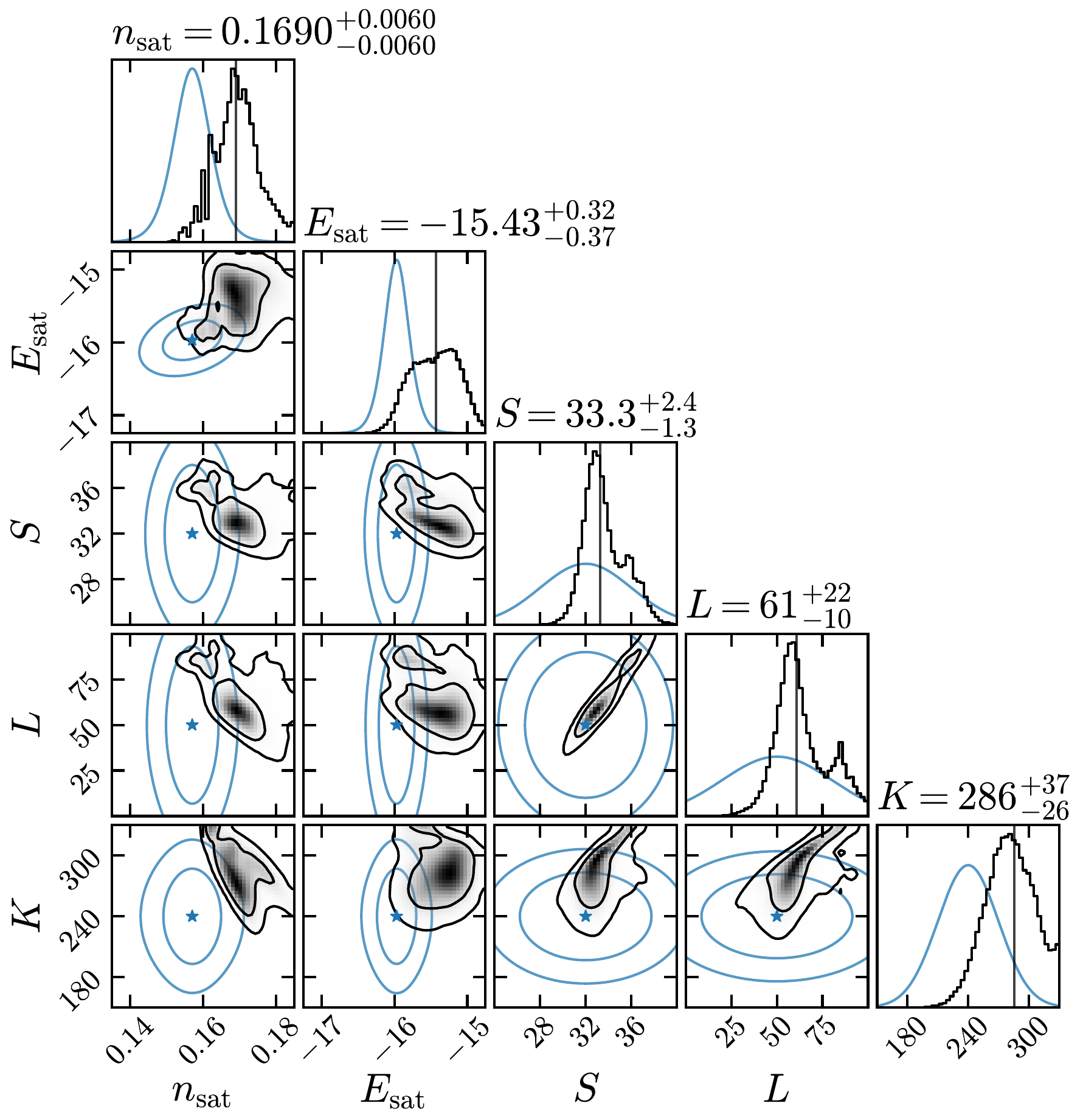}
    \caption{Corner plot showing the posterior distributions and pairwise correlations of the saturation properties $n_{\mathrm{sat}}$, $E_{\mathrm{sat}}$, $S$, $L$, and $K$ obtained from the \pmmimsrg{} Bayesian analysis.
    All observables are given in $\MeV$, except for $n_\mathrm{sat}$, which is given in $\fmiq$. The diagonal panels display the marginal posterior distributions with median values and credible intervals, while the off-diagonal panels show the corresponding joint posterior contours. The blue curves indicate the prior distributions: a bivariate student-$t$ distribution for $(n_{\mathrm{sat}}, E_{\mathrm{sat}})$ and independent Gaussian distributions for $(S, L, K)$, with the blue star marking the distributions' mean values. The $68\%$ and $95\%$ confidence-level contours are shown, and the reported uncertainties (centered on the median) are at the 68\% credibility level.}
    \label{fig:cornersat}
\end{figure}

\subsubsection{LEC constraints and predictions for the low-density EOS}

With the standard Monte Carlo sampler implemented in the Python package \texttt{emcee}~\cite{emcee}, we sample the posterior distribution of the LECs.
We use 32 random walkers, initialized at the maximum a posteriori (MAP) estimate plus
random perturbations, discard $\num{4000}$ burn-in steps, and generate $\num{20000}$ production steps.
The production chain was thinned by retaining every fifth sample.
Figure~\ref{fig:cornerlec} shows the resulting posterior samples of the LEC posterior distribution, which were subsequently propagated through the \pmmimsrg{} emulator to evaluate the posterior distributions of the saturation properties shown in Fig.~\ref{fig:cornersat}.
The titles report the 68\% confidence regions centered on the median value.

Figure~\ref{fig:cornerlec} shows that, particularly, $c_D = -2.8_{-1.6}^{+3.2}$ is not well constrained by this fit protocol, exhibiting long tails, and its distribution is cut off at the lower edge of the prior, $ |c_D| \leq 5$.
While positive values of $F_2$ are slightly favored, our constraints are statistically consistent with $F_2 = 0$ at the 68\% level.
We emphasize again that these NN and 3N interactions are constructed only for exploratory purposes, serving to investigate how \pmmimsrg{} can constrain nuclear forces to reproduce empirical saturation properties.

In Fig.~\ref{fig:cornersat}, the associated posterior for the saturation properties (black contours) remains consistent with empirical expectations (blue contours representing the prior), although the posterior consistently suggests higher values for those low-density EOS parameters, especially for $K \approx 286^{+37}_{-26} \, \MeV$.
In addition, we no longer observe a Coester-like anticorrelation of $n_{\mathrm{sat}}$ and $E_{\mathrm{sat}}$, as reflected by the Pearson correlation coefficient $\rho = 0.21$~\cite{Drischler:2024ebw}.
However, the mode of the predicted saturation point distribution, $p(n_\mathrm{sat},E_\mathrm{sat})$, is close to the 68\% contour of the empirical point, so closer than with $F_2 =0$ alone, as was expected.
These features reflect the fact that, in the present analysis, the LECs are constrained only by empirical saturation properties.
Consequently, the calibration does not yet incorporate important information from few-body observables, which is essential for further constraining the magnitudes of the LECs.
In future work, it will be interesting to incorporate constraints from finite nuclei, including the \isotope[3]{H} binding energy and \isotope[4]{He} charge radius, but also heavier nuclei, as was done in Ref.~\cite{Vernik:2025czp}.
Our \pmmimsrg{} emulator with conformal prediction can be applied straightforwardly to these nuclear structure calculations, providing a unified IMSRG framework for both finite nuclei and infinite matter.

Finally, we stress that this type of Bayesian calibration, based on our IMSRG calculations without \pmmimsrg{}, would have been computationally challenging, if not impossible, given the computational resources available to us.
Our emulator accelerates IMSRG calculations of the nuclear EOS by roughly five orders of magnitude, enabling efficient propagation of LEC uncertainties to the EOS and derived quantities, thereby enabling Bayesian workflows such as parameter estimation.

\subsection{{Propagating LEC uncertainties to the EOS}}
\label{subsec:uncertainty}

Next, we propagate $\num{10000}$ posterior LEC samples $(c_D, c_E, F_2)$, as depicted in Fig.~\ref{fig:cornerlec}, to the nuclear EOS to quantify its parametric uncertainties.
Specifically, using the \pmmimsrg{} emulator, we evaluate the energies of particles in PNM and SNM for $ n= [0.08,\,0.24]~\fmiq$ with a constant spacing of $0.01~\fmiq$.

In addition to these parametric uncertainties, for each LEC sample, we employ the pointwise Bayesian truncation error model developed by the BUQEYE collaboration~\cite{Melendez:2017phj,Melendez:2019izc} to estimate EFT truncation error in the energy per particle.
This truncation error arises from performing EFT calculations in practice at a finite chiral order.
Given a set of order-by-order many-body calculations in the chiral EFT expansion and physics-informed priors, the BUQEYE model learns the convergence pattern of these successive calculations and sums the contributions from omitted orders to all orders.
The pointwise model neglects important correlations in the density, which is an approximation that should be revisited.

We adopt $Q(k_F) = k_F/\Lambda_b$ as the EFT expansion parameter in medium, where $k_F(n) = 4 \pi\sqrt[3]{n /N}$ denotes the Fermi momentum in the finite box and $N$ the particle number (i.e., $N=66$ for PNM and $N=132$ for SNM), and take the breakdown scale to be $\Lambda_b = 600~\MeV$.
The model's hyperparameters are set to $\mu_0 = \tau_0 = 1$, and the reference scale is chosen to be $y_{\mathrm{ref}} = 16~\MeV$ for both PNM and SNM calculations.

\begin{figure}[tb]
    \centering
    \includegraphics[width=\linewidth]{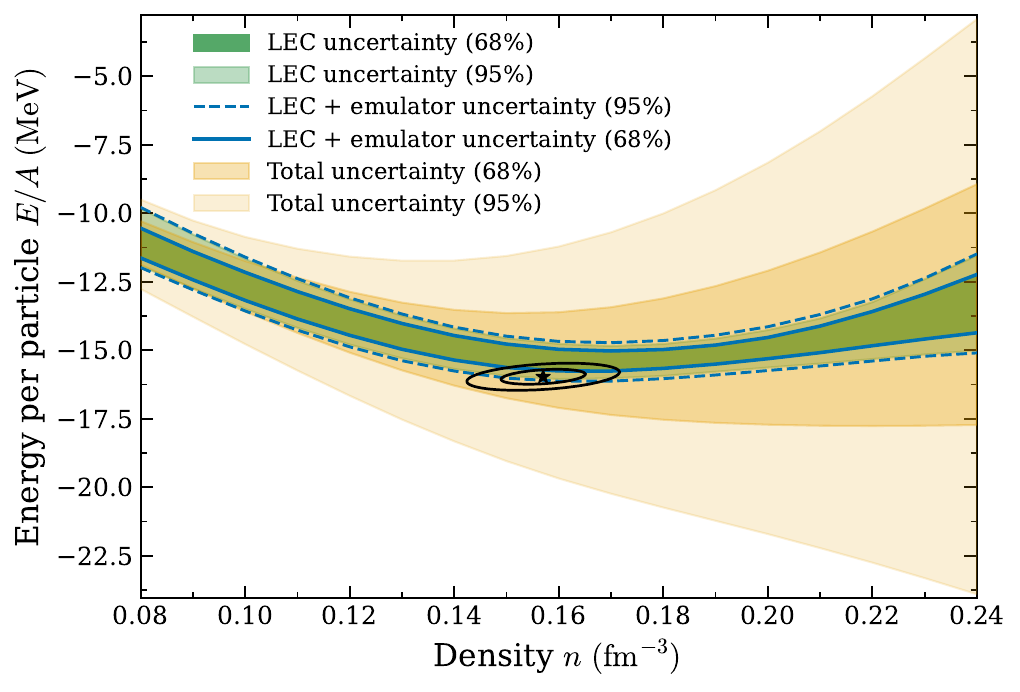}
    \caption{
    Same as Fig.~\ref{fig:snm_cd_ce_variationmc} but for the LEC samples in Fig.~\ref{fig:cornerlec}.
    }
    \label{fig:pmmsnmuqmc_post}
\end{figure}

\begin{figure}[tb]
    \centering
    \includegraphics[width=0.999\linewidth]{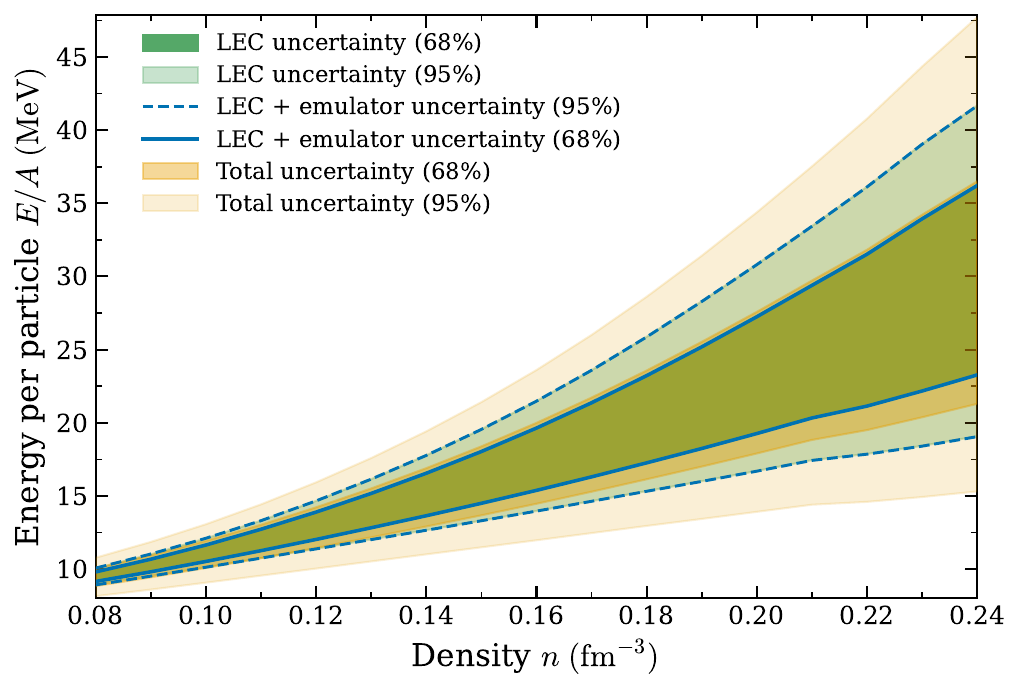}
    \caption{Same as Fig.~\ref{fig:pmmsnmuqmc_post} but for PNM, where $c_D$ and $c_E$ do not contribute in contrast to $F_2$.}
    \label{fig:pmmpnmuqmc_post}
\end{figure}

With these prerequisites in place, we construct credibility intervals for our IMSRG calculations.
Figures~\ref{fig:pmmsnmuqmc_post} and~\ref{fig:pmmpnmuqmc_post} show the corresponding uncertainty bands in the SNM and PNM EOSs, respectively, obtained using the \pmmimsrg{} emulator.
The green bands correspond to the LEC uncertainty; the blue solid and dashed curves delimit the combined LEC and emulator uncertainties at the 68\% and 95\% credibility levels, respectively; the orange bands represent the total uncertainty, including the EFT truncation error, at the corresponding credibility levels.
In SNM, as one might have expected, we observe a clear hierarchy of uncertainties, with the EFT truncation error as the dominant source and the emulator uncertainties the smallest, indicating that omitting EFT truncation errors from the likelihood is an important limitation of the present Bayesian calibration and should be revisited in the future.
In PNM, we find that the propagated LEC uncertainty band is wider than in SNM at densities above $0.09~\fmiq$, and the hierarchy of uncertainties found in SNM is no longer observed.
Although this may appear counterintuitive at first, it is a direct consequence of the applied fit protocol.
The PNM EOS is mainly constrained by the weakly informed prior on the symmetry energy and its slope parameter evaluated at the saturation density.
The latter is proportional to the pressure in PNM at that density.
Since the saturation point in SNM is more tightly constrained and $F_2$ contributes to the PNM EOS, whereas $c_D$ and $c_E$ do not, we obtain only weakly constrained results for the PNM EOS.
It would be interesting to incorporate additional PNM constraints, e.g., from nuclear experiments and neutron star observations, into this fit protocol.

\begin{figure}
\includegraphics[width=0.999\linewidth]{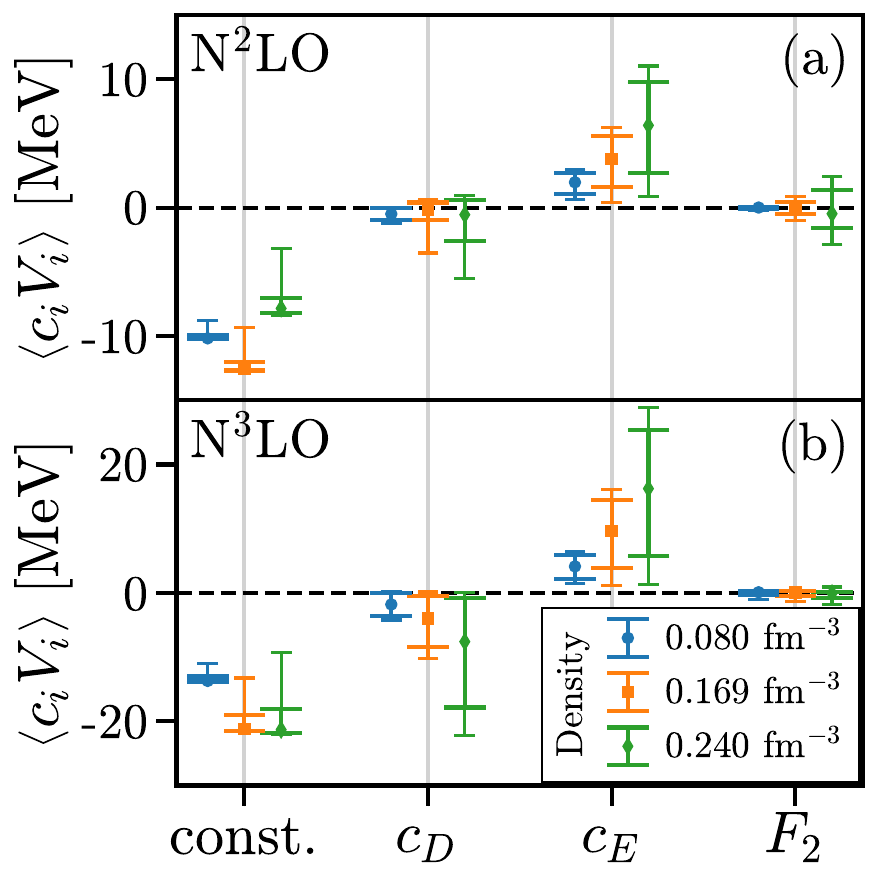}
\caption{Median expectation values of each term in the emulated Hamiltonian with $68\%$ and $95\%$ credible intervals from posterior LEC samples at N$^2$LO (panel a) and N$^3$LO (panel b) at three different densities for SNM. The labels on the horizontal axis indicate the individual $c_iV_i$ term where ``$\text{const.}$'' denotes the fixed $V_0$ term in the emulator.}
\label{fig:exp}
\end{figure}

In Ref.~\cite{Vernik:2025czp}, the impact of including $F_2$ was analyzed for light- and medium-mass nuclei.
They found that while its inclusion can have a significant impact on the ground state energies and charge radii, this is mainly due to changes in the other short-range 3N couplings, $c_D$ and $c_E$.
In particular, they found that the expectation value of the $F_2$ term was much smaller than the $c_D$ and $c_E$ terms.
Since our emulator internally mimics the form of the Hamiltonian for the EMN interaction via Eq.~\eqref{eq:freespaceH}, this allows us to extract the individual operators from the PMM, along with the emulated ground state wavevector, to compute the expectation values of the $F_2$, $c_D$, and $c_E$ terms, as well as the remaining part of the Hamiltonian, denoted by $V_0$, that does not depend on these 3N LECs in the ground state. Figure~\ref{fig:exp} shows these expectation values and the associated $68\%$ and $95\%$ credible intervals obtained from posterior LEC samples at N$^2$LO (panel a) and N$^3$LO (panel b), at the posterior median saturation density $n_{\rm sat}= 0.169~\fmiq$, and the minimum and maximum densities considered in the training data, $n_\text{min}=0.08~\fmiq$ and $n_{\rm max}= 0.240~\fmiq$.
Consistent with the findings of Ref.~\cite{Vernik:2025czp}, the expectation value of the $F_2$ term in all cases is significantly smaller in magnitude than the $c_D$, $c_E$, and $V_0$ terms at all three densities.

Taken together, the results of our exploratory studies of $F_2$ contributions demonstrate the viability of the present emulator-driven Bayesian framework and highlight an important limitation.
Additional constraints from light (and even medium-mass) nuclei would help improve the predictive power of these chiral NN and 3N forces for nuclear structure and nuclear matter calculations.
Adding these complementary constraints to the Bayesian fit protocol will be an important direction for future work~\cite{Vernik:2025czp}.

\section{Summary and Outlook}
\label{sec:summary}

We have developed a fast, accurate, and physically-consistent PMM emulator for (nonperturbative) IMSRG calculations of the nuclear EOS~\cite{Udiani_thesis,Udiani_IMSRG} in PNM and SNM across the space of LECs, density, flow parameter, and basis size.
This emulator enables principled UQ of the nuclear EOS, including emulator error estimation and LEC variations, a task that was previously computationally prohibitive.
The emulator uncertainty estimation via conformal predictions developed in this work is generally applicable and not limited to EOS or IMSRG calculations.
It also enables controlled extrapolation of slowly converging (or even diverging) IMSRG calculations to much lower resolution scales.
The resulting low-resolution potentials could be used to accelerate the convergence of many-body approaches based on diagrammatic expansions of the many-body Schr{\"o}dinger equation, such as MBPT, thereby facilitating \textit{ab~initio} studies of low-density nuclear matter across a wide range of many-body frameworks.

To demonstrate the efficacy of this framework, we propagated parametric uncertainties of order-by-order chiral NN and 3N interactions up to \nthlo{}~developed in Ref.~\cite{Drischler:2017wtt} to the EOS in the limits of PNM and SNM.
Leveraging our \pmmimsrg\ emulator, we achieved computational speedups of five orders of magnitude, emphasizing that these calculations would not have been feasible (given our computational resources) without emulation.
Together with quantified emulator and EFT truncation errors, this enabled principled UQ that captured the dominant sources of theoretical uncertainty in the nuclear EOS.

We then constructed exploratory chiral NN and 3N interactions that incorporate the promoted quark-mass-dependent 3N interaction governed by $F_2$ (with the other proportional to $D_2$ omitted, following Ref.~\cite{Vernik:2025czp}), and calibrated the leading 3N LECs $c_D$, $c_E$, and $F_2$ to empirical saturation properties of nuclear matter using Bayesian parameter estimation.
These interactions are not intended to be state-of-the-art, but rather to serve as explorations of constraining nuclear forces using the IMSRG to reproduce empirical saturation properties.
Taken together, our results demonstrate the viability and efficacy of the
\pmmimsrg\ framework for Bayesian calibration and uncertainty quantification, while highlighting the need for complementary constraints from few-body systems and finite nuclei to pin down the allowed values and the impact of these LECs.

Looking ahead, we plan to extend this framework to emulate the EOS across a range of proton fractions and temperatures, enabling the construction of uncertainty-quantified nuclear EOSs derived from microscopic Hamiltonians and directly applicable to astrophysical simulations of supernovae and binary mergers.
Furthermore, it will be of interest to perform a joint analysis of light- to medium-mass nuclei and nuclear matter within our \pmmimsrg\ emulator framework, and to systematically combine EFT truncation
and emulator uncertainties in the Bayesian calibration, which can be straightforwardly extended to these nuclear-structure calculations.
This effort can be complemented by recent Bayesian analysis of two- and three-body scattering observables facilitated by fast and accurate scattering emulators~\cite{Gnech:2025gsy,Gnech:2025lbg,Maldonado:2025ftg,Giri:2025pkw,Heihoff:2026ycq}.
Such a combined analysis has the potential to shed light on the importance of nuclear saturation properties in infinite matter for reproducing binding energies and charge radii of medium-mass to heavy nuclei~\cite{Hoppe:2019uyw, Hu:2025cjl}.
Altogether, these advances in emulator-driven \textit{ab~initio} calculations, with theoretical uncertainties rigorously quantified, are highly exciting, as they pave the way for new insights into strongly interacting, strongly correlated systems that appeared computationally infeasible until recently.

\section*{Acknowledgements}
We thank V.~Cirigliano, M.~Dawid, W.~Dekens, M.~Kumamoto, and S.~Reddy for useful discussions.
We are also grateful to our DOE STREAMLINE2 collaborators for guidance and encouragement.
K.Y.\ and S.B.\ were supported in part by the National Science Foundation (NSF) Grants PHY-2013047 and PHY-2310020. P.C.\ was supported in part through the Department of Energy award DE-SC0026198.
C.D.\ was supported by the NSF under award PHY-2339043 and the U.S.\ Department of Energy, Office of Science, Office of Nuclear Physics, under the FRIB Theory Alliance award DE-SC0013617. This work was supported in part through computational resources and services provided by the Institute for Cyber-Enabled Research (ICER) at Michigan State University.
The following Python libraries were used to generate the results and visualizations in this work:
\texttt{corner}~\cite{corner},
\texttt{emcee}~\cite{emcee},
\texttt{JAX}~\cite{jax2018github},
\texttt{Jupyter}~\cite{jupyter},
\texttt{Matplotlib}~\cite{Hunter:2007},
\texttt{NumPy}~\cite{harris2020array},
\texttt{pyPMM}~\cite{pyPMM2026}, and
\texttt{SciPy}~\cite{2020SciPy-NMeth}.

\section*{Data Availability}
The data that support the findings of this article are available upon reasonable request from the authors and will be publicly available at the time of publication.

\medskip

\bibliographystyle{apsrev4-1}
\bibliography{refs,bayesian_refs,bib}

\end{document}